\documentclass{aa}  

\usepackage{txfonts}
\usepackage{multirow}
\usepackage{subfig,graphicx}
\usepackage{float}
\usepackage{natbib,twoopt}
\usepackage[breaklinks=true]{hyperref}
\hypersetup{
    colorlinks=true,
    linkcolor=blue,
    filecolor=blue,      
    urlcolor=blue,
    citecolor=blue,
}
\bibpunct{(}{)}{;}{a}{}{,}            

\begin{document} 

   \title{Exploring the physical properties of Type II Quasar candidates at intermediate redshifts with \texttt{CIGALE}}
    \titlerunning{Exploring the physical properties of QSO2 candidates at intermediate redshifts}
   \author{P.~A.~C. Cunha\inst{1,2} \fnmsep\thanks{Corresponding author, \email{pedro.cunha@astro.up.pt}}
          \and
          A. Humphrey\inst{2,3}
          \and
          J. Brinchmann\inst{1,2}
          \and
          A. Paulino-Afonso\inst{2}
          \and
          L. Bisigello\inst{4,5}
          \and
          M. Bolzonella\inst{6}
          \and
          D. Vaz\inst{1,2}
          }

    \institute{Departamento de Física e Astronomia, Faculdade de Ciências, Universidade do Porto, Rua do Campo Alegre 687, PT4169-007 Porto, Portugal
         \and
             Instituto de Astrofísica e Ciências do Espaço, Universidade do Porto, CAUP, Rua das Estrelas, PT4150-762 Porto, Portugal
        \and 
            DTx – Digital Transformation CoLab, Building 1, Azurém Campus, University of Minho, PT4800-058 Guimarães, Portugal
        \and
            INAF–Osservatorio Astronomico di Padova, Vicolo dell’Osservatorio 5, I-35122, Padova, Italy
        \and
            Dipartimento di Fisica e Astronomia "G. Galilei", Università di Padova, Via Marzolo 8, 35131 Padova, Italy
        \and
            INAF-Osservatorio di Astrofisica e Scienza dello Spazio di Bologna, Via Piero Gobetti 93/3, 40129 Bologna, Italy
             }

   \date{Received ; accepted }

\abstract{Active Galactic Nuclei (AGN) play a vital role in the evolution of galaxies over cosmic time, significantly influencing their star formation and growth. As obscured AGNs are difficult to identify due to obscuration by gas and dust, our understanding of their full impact is still under study. Investigating their properties and distribution, in particular Type II quasars (QSO2), is essential to comprehensively account for AGN populations and understand how their fraction evolves over time. Such studies are important to provide critical insights into the co-evolution of AGNs and their host galaxies.}
{Following our previous study, where a machine learning approach was applied to identify 366 QSO2 candidates from SDSS and WISE surveys (median \(z \sim 1.1\)), we now aim to characterise this QSO2 candidate sample by analysing their spectral energy distributions (SEDs) and derive their physical properties.}
{We estimated relevant physical properties of the QSO2 candidates, including star formation rate (SFR), stellar mass (\(M_\ast\)), AGN luminosity, and AGN fraction, using SED fitting with \texttt{CIGALE}. We compared the inferred properties with analogous populations in the semi-empirical simulation \texttt{SPRITZ}, placing these results in the context of galaxy evolution.}
{The physical properties derived for our QSO2 candidates indicate a diverse population of AGNs at various stages of evolution. In the SFR-\(M_\ast\) diagram, QSO2 candidates cover a wide range, with numerous high-SFR sources lying above the main sequence at their redshift, suggesting a link between AGN activity and enhanced star formation. Additionally, we identify a population of apparently quenched galaxies, which may be due to obscured star formation or AGN feedback. Furthermore, the physical parameters of our sample align closely with those of composite systems and AGN2 from \texttt{SPRITZ}, supporting the classification of these candidates as obscured AGNs.}
{This study confirms that our QSO2 candidates, selected via a machine learning approach, exhibit properties consistent with being AGN-host galaxies. This method proves to be reliable at identifying AGNs within large galaxy samples by considering AGN fractions and their contributions to the infrared luminosity, going beyond the limitations of traditional colour-colour selection techniques. The diverse properties of our candidates demonstrate the capability of this approach to identify complex AGN-host systems that might otherwise be missed.  This shows the help that machine learning can provide in refining AGN classifications and advancing our understanding of galaxy evolution driven by AGN activity with new target selection.}


   \maketitle
%


\section{Introduction}

Galaxies and active galactic nuclei (AGN) have an interlinked relationship, each being potentially able to affect the evolution of the other. An AGN consists of a supermassive black hole (SMBH; $10^{6-9}$M$_{\odot}$) with an accretion disk made of dust and gas. As interstellar matter falls under the gravitational influence of the SMBH, the accretion process produces a highly energetic signature observable across the entire electromagnetic spectrum \citep[e.g.,][]{1943ApJ....97...28S,1969Natur.223..690L}. AGN feedback, positive or negative, can directly influence galaxy evolution by quenching or enhancing star formation \citep[e.g.,][]{2012ARA&A..50..455F,2015ApJ...810...74A}. Therefore, the evolution of SMBHs and their host galaxies are intrinsically linked, exemplified by AGN feedback, SMBH growth, and the M-$\sigma$ relation \citep[e.g.,][]{1998Natur.395A..14R,1998AJ....115.2285M,2000ApJ...543L...5G,2001MNRAS.320L..30M,2001ApJ...547..140M,2005SSRv..116..523F,2010MNRAS.402.2453D,2012NewAR..56...93A,2013ARA&A..51..511K,2018NatAs...2..198H,2023A&A...672A.137L}.

AGN unification schemes propose a paradigm where inclination, i.e., the line of sight angle, is responsible for some of the differences in the observed properties of AGNs \citep{1993ARA&A..31..473A, 1995PASP..107..803U}. Among the simplest AGN unification schemes is the separation of sources into two classes based on the orientation of an obscuring dusty torus with respect to our line of sight: Type I, called unobscured AGNs; and Type II, called obscured AGNs (AGN2). Type II quasars (QSO2) are luminous AGN2s with a dusty torus, oriented to hide the SMBH accretion disk from our line of sight \citep[e.g.,][see references therein for a more complete review]{2017A&ARv..25....2P,2018ARA&A..56..625H}. Recent infrared (IR) and submillimeter observations introduce a new scenario in which a two-component dusty structure with equatorial and polar features is present, and radiation pressure plays an important role \citep[][and references therein]{2019ApJ...884..171H,2023MNRAS.519.3237S}.

The obscuration of the accretion disk at ultraviolet and optical wavelengths acts as a 'natural coronagraph,' enabling the study of galaxy and AGN co-evolution and the physical properties of the host galaxies, which would be challenging with the often overwhelming glare of the optically luminous accretion disk \citep[e.g.,][]{2009A&A...507.1277B,2011A&A...534A.110L,2013MNRAS.433.1015S,2017A&A...602A.123L}. Therefore, AGN2 are useful for understanding the evolution of galaxies across cosmic time \citep[e.g.,][]{1998AJ....115.2285M,2008ApJ...676...33D,2011ApJ...732....9G,2012MNRAS.426..276B,2014MNRAS.438.1839B,2015MNRAS.454.4452H,2017A&ARv..25....2P,2020A&A...634A.116V,2021A&A...650A..84V}. 

Multiwavelength studies, particularly in the X-ray and IR ranges, are crucial for identifying and studying AGN2 and QSO2 \citep[e.g.,][]{2006ApJ...642...81S,2009cfdd.confE.147M,2009MNRAS.400.1199R,2014A&A...565A..19R,2016agnt.confE..68V,2020ApJ...897..160L,2021ApJ...908..185C,2023MNRAS.521..818S,2023ApJ...951...27Y}. While X-ray observations are highly effective for detecting AGNs and estimating their obscuration \citep[e.g.,][]{2005ARA&A..43..827B,2011ApJS..195...10X,2017ApJS..232....8L}, heavily obscured sources can sometimes be missed \citep[e.g.,][]{2011A&A...526L...9C,2012ApJ...748..142D,2021AJ....162...65H,2021ApJ...908..185C,2023ApJ...950..127C}. Despite this, optical to near-IR (NIR) observations remain the primary method for selecting AGN2 due to the abundance of data \citep[e.g.,][]{2007ApJ...671.1365H,2016ApJ...818...88L}. These recent studies show the ongoing importance of multiwavelength observations in the study of the complexity of the obscured AGN population.

Whenever spectroscopic information is missing to study and derive the physical properties, photometric information can provide a useful alternative. By taking advantage of different types of spectral energy distribution (SED) models, i.e., empirical, semi-empirical, and theoretical, SED fitting tools have shown to be a reliable method to estimate photometric redshifts or to constrain physical properties \citep[e.g.,][]{2000A&A...363..476B,2003MNRAS.338..733B,2007ApJ...663...81P,2009ApJ...690.1250S,2013ARA&A..51..393C,2023ApJ...944..141P}. However, powerful tools also have shortcomings or caveats. For example, SED fitting is very dependent on initial assumptions and is sensitive to the quality and diversity of the photometric data. These factors can introduce uncertainties and biases in the computed properties, requiring careful interpretation and, where possible, complementary data to validate results \citep[e.g.,][]{2022A&A...662A..86P}.

The selection of QSO2 sources, in the optical, has relied on a combination of a few emission line properties or subjective qualitative criteria \citep{2003AJ....126.2125Z,2004AJ....128.1002Z,2013MNRAS.435.3306A,2014AAS...22311504R}. Although this kind of methodology is extremely valuable for the identification and characterisation of QSO2s, it requires optical spectroscopic information and specific emission lines to be detected or present in the spectrum (e.g. CIV$\lambda1549$; [OIII]$\lambda\lambda4959,5008$). 

In \citet{2024A&A...687A.269C}, hereafter C2024, we presented \texttt{AMELIA}, a novel machine learning pipeline that incorporates few-shot learning and generalised stacking to identify QSO2 candidates using optical and IR photometry. In C2024, we used the candidates identified by \citet{2013MNRAS.435.3306A}, with a focus on developing a quick and efficient methodology to identify "hidden" QSO2 sources in large surveys. We used a combination of photometric magnitudes and optical, IR and optical-IR colours as input into the machine learning pipeline allowing it to learn multiple relationships, bypassing the limitations of typically lossy colour-colour selection criteria. 

We identified a sample of 366 QSO2 candidates within the redshift desert ($1 \leq z \leq 2$).The SED fitting code \texttt{CIGALE} was used to estimate the contribution of AGN to the total IR dust luminosity for the QSO2 candidate sample. Here, we perform a more comprehensive SED fitting to estimate the host galaxies' physical properties and contextualise the sources into the overall process of galaxy evolution.

This paper is organised as follows. Section \ref{section:Data} details our photometric data selection and QSO2 candidate selection methodology. In Section \ref{section:sed_fitting}, we describe the SED fitting tools used in this work, along with their setup. Section \ref{section:results_lephare} tests the reliability of photometric redshift estimation using the SED fitting tool \texttt{LePhare++} with spectroscopic redshifts from SDSS. Section \ref{section:results_cigale} provides analysis and estimations of the physical properties of the QSO2 candidates using \texttt{CIGALE}. In Section \ref{section:multiwavelength}, we explore the addition of radio and X-ray photometry to the SED fitting and how these influence the derived physical properties. Section \ref{section:SFR_separation} investigates the use of the derived star-formation rate (SFR) to gain insights into their physical nature. Section \ref{section:comparison_literature} compares our results with the current literature. In Section \ref{section:SPRITZ}, we perform a comparison with the simulated semi-empirical catalogue \texttt{SPRITZ}. Finally, Section \ref{section:conclusions} summarises our conclusions. Throughout this paper, we adopt a flat-universe cosmology with \(H_0 = 69.3\) km s\(^{-1}\) Mpc\(^{-1}\) and \(\Omega_M = 0.286\) \citep[WMAP 9-year results,][]{2013ApJS..208...19H}, similar to \citet{2022ApJ...927..192Y}.

\section{Photometric data}
\label{section:Data}

In this work, we use data from the Sloan Digital Sky Survey \citep[SDSS;][]{SDSS16,2022ApJS..259...35A}, Wide-field Infrared Survey Explorer \citep[WISE;][]{2010AJ....140.1868W}, X-ray Multi-Mirror Mission \citep[XMM-Newton;][]{2012ApJ...756...27L}, and the LOw-Frequency ARray (LOFAR) Two-metre Sky Survey \citep[LoTSS;][]{2022A&A...659A...1S}. In the following subsections, we describe the different samples used in this study and their motivation.

\subsection{Control sample and QSO2 candidates}
\label{section:Data_qso2}
From the SDSS DR16 spectroscopic galaxy sample, we selected sources based on photometric constraints that meet the following criteria (AB magnitudes): (i) \( 19 \leq u \leq 26 \); (ii) \( 19 \leq g \leq 24 \); (iii) \( 19 \leq r \leq 24 \); (iv) \( 19 \leq i \leq 24 \); (v) \( 19 \leq z \leq 25 \). This selection resulted in approximately 22,000 galaxies for identifying QSO2 candidates (see the methodology in Section 3 in C2024). These criteria were chosen to emulate the optical observational constraints applied to the class A sample described in \citet{2013MNRAS.435.3306A}, where QSO2 objects are identified by narrow emission lines (e.g., Ly\(\alpha \lambda 1216\) and CIV\(\lambda1549\)), weak continuum, absence of associated absorption features, and high equivalent width (EW), within a redshift range of \(2 \leq z \leq 4\). For further details on how this sample was used to derive QSO2 candidates, see C2024 \citep{2024A&A...687A.269C}.

The selected galaxies were then used to identify QSO2 candidates using \texttt{AMELIA}. From this sample, we selected the candidates assuming a QSO2 classification probability $\geq 0.8$. As a final product of our pipeline, we obtained 366 QSO2 candidates, with redshift between 1 and 2, the so-called redshift desert. In C2024, we added further observational evidence for the AGN nature of the QSO2 candidates by comparing photometric colour-colour criteria, performing spectroscopic analysis whenever possible, and estimating the fraction of AGN using \texttt{CIGALE}.

From the sample of non-QSO2 candidates, i.e. those with a QSO2 classification probability $\leq 0.5$, we randomly extracted 366 galaxies to serve as a control sample. This control sample is expected to reveal distinct properties in the colour-colour photometric space across both optical and IR wavelengths. By comparing their estimated physical properties with those of QSO2 candidates, we can validate the refined selection criteria provided by \texttt{AMELIA}.

\subsection{Bongiorno sample}

In \citet{2012MNRAS.427.3103B}, obscured AGNs were selected using a [NeV]$\lambda3426$-based optical selection, within the zCOSMOS bright survey\footnote{\url{https://cdsarc.cds.unistra.fr/viz-bin/cat/J/MNRAS/427/3103}} \citep{2007ApJS..172...70L}. The catalogue presents four classifications derived using spectra, \texttt{Classsp}: class 1, unobscured AGNs; class 2, obscured AGNs classified from the X-ray; class 22.2, obscured AGNs selected using the BPT diagram; class 22.4, obscured AGNs selected through the [NeV] emission line. 

Two selection constraints were applied to the Bongiorno sample: $1 \leq z \leq 2$; and \texttt{Classsp} $\neq$ 1. The first constraint ensures that the sample is within the redshift range of our candidate sample. The second removes the unobscured AGN from the sample. The final sample has 87 sources, with 78 sources with class 2 and 9 sources with class 22.4. In the catalogue, the authors also provide stellar mass and SFR estimates derived from photometric SED fitting. This sample will serve as an AGN control group, allowing us to compare the physical properties of our QSO2 candidates.

\section{SED fitting}
\label{section:sed_fitting}

While the use of semi-supervised machine learning-based methodologies can help to accelerate astronomical analysis, it is crucial to validate and analyse the output of such methods. Since our QSO2 candidates do not have high S/N spectra available, we will base our inference of physical parameters on the photometric data available. When SED fitting techniques are explored, it can easily transform into a very complex and fascinating problem. In this work, we use the state of the art SED implementations to characterise our candidate sample, while taking into consideration the caveats of only using optical and IR data from SDSS and WISE.   

\subsection{\texttt{LePhare++}}

\texttt{LePhare++}\citep{1999MNRAS.310..540A,2006A&A...457..841I,2011ascl.soft08009A} \footnote{\url{https://github.com/lephare-photoz/lephare/releases/tag/v0.1.13}} is an SED fitting code optimised to compute photometric redshifts. It uses a chi-square minimisation approach to compare observed photometric data with theoretical models or SED templates. For each redshift, the code generates model photometry by redshifting the SED templates and integrating them over the filter transmission curves. Through a chi-square minimisation, \texttt{LePhare++} identifies the best-fit class and redshift, for the observed photometry, and takes into account photometric errors in its calculations \citep[e.g.][]{2022MNRAS.513.3719H}. In this work, \texttt{LePhare++} will be used to compare photometric redshifts with SDSS spectroscopic redshifts. 

In our usage of \texttt{LePhare++}, we will take into consideration two outputs based on the chi-square minimisation: best-fit class and redshift. We consider libraries for stars, galaxies and QSOs. In particular, we are interested in the templates for composite systems, as we expect them to be the more suitable solution for our sample. For this study, we considered a comprehensive set of spectral templates tailored to capture the diversity in galaxy morphologies, AGN contributions, and stellar populations. The templates included a range of hybrid galaxy-AGN models from \citet{2009ApJ...690.1250S}, with various percentages of AGN contribution, allowing for the modelling of composite galaxies. Additionally, we included templates from \citet{2009ApJ...690.1236I}. Furthermore, a range of stellar templates from \citet{1998PASP..110..863P} were used to ensure accurate fitting across spectral types. This diverse template set allows the SED fitting to account for both galaxy and AGN contributions across a wide range of redshifts and physical conditions. We allow the redshift to vary between 0 and 6, with constants steps $\delta = 0.1$.

\subsection{\texttt{CIGALE}}
\label{section:cigale_setup}

To better characterise our sample, we required a more sophisticated SED fitting code capable of inferring both the physical properties of the host galaxy and the AGN within it. We used the \texttt{Code Investigating GALaxy Emission} \citep[\texttt{CIGALE};][, version 2022.1]{2005MNRAS.360.1413B, 2009A&A...507.1793N,2019A&A...622A.103B,2022ApJ...927..192Y}. \texttt{CIGALE} performs the SED fitting within a self-consistent framework that balances UV/optical absorption with IR emission. By analysing the best-fit model, it can estimate various parameters, including the contributions from stellar, AGN, and star formation components, to match the photometric data from the rest-frame UV to the far-infrared (FIR) bands.

To model the star formation histories (SFHs) of our target galaxies, we adopted a delayed SFH with an optional exponential burst (\texttt{sfhdelayed}) within \texttt{CIGALE}. This model provides a good balance between simplicity and the ability to capture the key features expected in the SFHs of our sample, which includes both early-type and late-type galaxies. Mathematically, this SFH is represented as:

\begin{equation}
    \rm{SFR}(t) \propto \frac{t}{\tau ^2} \times exp(-t/\tau), t \in [0,t_0],
\end{equation}

where $t_0$ represents the age of the galaxy, i.e. the time since the onset of star formation, and $\tau$  corresponds to the time at which the star formation rate (SFR) reaches its peak. By adjusting the parameter $\tau$, we can effectively model a range of SFHs, from those with a rapid rise and fall in star formation, small $\tau$, to those with a more gradual evolution, large $\tau$.

This SFH model assumes that star formation begins after a delay, allowing for a period of quiescence before the onset of significant star formation activity \citep{2018A&A...620A..50M}. The optional exponential burst component enables us to account for potential short-lived increases in star formation that may occur.

For the stellar populations, we used the \citet{2003MNRAS.344.1000B} single stellar population models with the \citet{2003PASP..115..763C} initial mass function.  We assumed solar metallicity (Z=0.02) for all SSPs. To account for the effects of dust on the observed light, we adopted the models from \citet{2000ApJ...533..682C} and \citet{2002ApJ...574..114L}. This approach allows for differential reddening between young and old stellar populations, which we separated by an age of 10 Myr. This separation is motivated by the assumption that very young stars are still embedded in their dust-producing birth clouds, whereas older stars have migrated away from these regions.  The colour excess of the nebular emission lines, E(B-V), was allowed to vary between 0.1 and 0.9, with a step of 0.1, to explore a range of dust attenuation levels.

Finally, to model the infrared emission from dust, we used the model from \citet{2014ApJ...784...83D}. This model considers the dust emission as a function of the radiation field intensity (U), with a single parameter, $\alpha$, controlling the distribution of the dust mass with respect to U. 

To account for the potential contribution of an active galactic nucleus (AGN) to the overall SED, we included the \texttt{SKIRTOR} model \citep{2012toru.work..170S, 2016MNRAS.458.2288S}. This model provides a sophisticated representation of the AGN torus by considering it to be composed of dusty clumps. Specifically, the \texttt{SKIRTOR} model assumes that 97\% of the torus mass is distributed in high-density clumps, while the remaining 3\% is smoothly distributed. This two-phase distribution allows for a more realistic representation of the torus density structure. Additionally, the model accounts for the anisotropic emission from the accretion disk, which is the flattened disk of material spiralling onto the black hole.

To explore a range of possible AGN geometries, we considered four different values for the opening angle ($oa$) of the torus: 20, 40, 60, and 80 degrees. The opening angle is measured between the equatorial plane of the torus and its outer edge. The half-opening angle of the dust-free region, which represents a clear line of sight towards the central engine, is then defined as 90°-$oa$. We also varied the inclination parameter, which is the angle between our line of sight and the AGN axis, from 30° to 90°, in steps of 10°. This allows us to account for different viewing angles of the AGN.

For the extinction of light passing through the polar regions of the torus, we adopted the extinction law of polar dust from \citet{2004ApJ...616..147G}. The amount of extinction in the polar direction, quantified by the colour excess E(B-V), was allowed to vary between 0 and 1 magnitude (see Table \ref{tab:SED_fitting_params_QSO2}). 

To quantify the relative contribution of the AGN to the total infrared (IR) dust luminosity, we used the AGN fraction, frac$_{\mathrm{AGN}}$, defined as:

\begin{equation}
    \mathrm{frac_{AGN}} = \mathrm{\frac{L_{dust, AGN}}{L_{dust, AGN} + L_{dust, galaxy}}}
\end{equation}

where L$_{\rm{dust, AGN}}$ is the luminosity of the dust heated by the AGN, and L$_{\rm{dust, galaxy}}$ is the luminosity of the dust heated by stars within the galaxy.  Both luminosities are integrated over all wavelengths. We allowed frac$_{\mathrm{AGN}}$ to vary from 0 to 0.99, covering a wide range of AGN contributions, from negligible ($\sim 0$) to dominant ($\sim 1$).


\section{Testing the SDSS spectroscopic redshifts}
\label{section:results_lephare}

\texttt{CIGALE} requires the redshift to be fixed during the fitting process. Thus it is important to validate the spectroscopic redshifts from SDSS, since they are in the redshift desert. While SDSS redshifts in general are reliable, our sample is extreme and shows very noisy spectra, sometimes with clearly wrong redshifts \citep[e.g.][see discussion in Sect. 2.3 for QSO2 sources at z \textless 1]{2016MNRAS.462.1603Y}, and in some cases we are unable to see any emission lines, in sources analogous to the 'line-dark' radio galaxies \citep[][see also Sect. 8 in C2024]{2015MNRAS.447.3322H,2016A&A...585A..32H}. To assess this reliability, we took two different approaches: spectroscopic validation via spectral stacking; and photometric redshift estimation using \texttt{LePhare++} SED fitting code.

\subsection{Spectral stacking}

To apply the spectral stacking technique, we collected the available spectra for the QSO2 sample from SDSS DR17. In total, we obtained 310 spectra, no flags or warnings were considered here, with spectroscopic redshifts derived using \texttt{idlspec2d}. Given the lack of high signal-to-noise ratio spectroscopic data, we did not correct for possible contaminants in the sample when producing the final stacked spectrum. Although some contamination is anticipated due to the sample's non-pure composition, this does not interfere with one of our main objectives, validating the AGN nature of the sources. Each spectrum was then moved to the rest frame wavelength, normalised by the mean flux, and then summed into a single spectrum after spline-interpolation.

For sources with an identifiable emission line, e.g. [OII]$\lambda\lambda3727,3730$, if the derived spectroscopic redshift is incorrect, or with a high deviation from the true redshift, we expect the stacked spectrum to have multiple features to appear at unexpected wavelengths. Moreover, small redshift deviations will only produce an artificial broadening effect in the detected emission lines. In contrast, noisy or featureless spectra will be cancelled out during the stacking procedure.

\begin{figure*}
    \centering
    \includegraphics[width=0.8\linewidth]{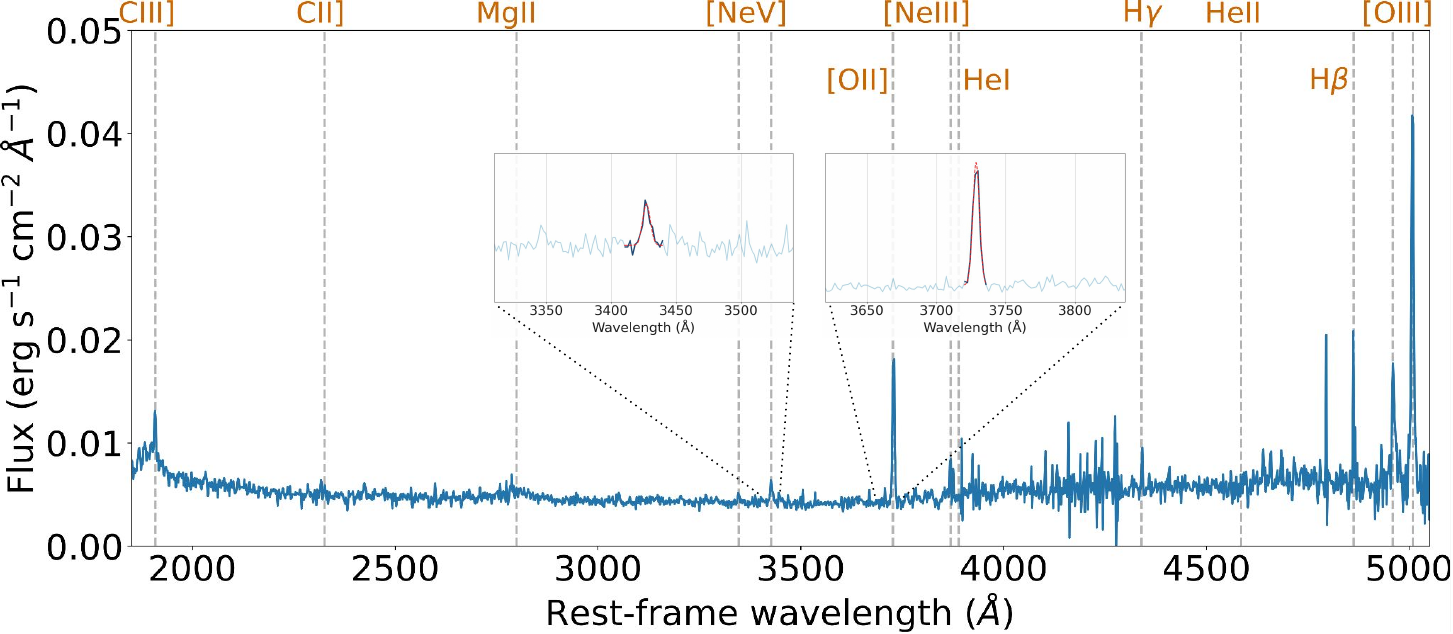}
    \caption{Stacked spectrum of QSO2 candidates with normalized median flux in the rest-frame wavelength, obtained through spline interpolation using available SDSS spectra. The spectrum is additionally normalized based on the number of contributing spectra for each rest-frame wavelength. Grey vertical lines mark the positions of emission lines in the rest frame. The plot includes two zoomed-in regions: (1) 3326–3526 $\AA$, highlighting the [NeV]$\lambda3426$ emission line, and (2) 3630–3830 $\AA$, highlighting the [OII]$\lambda\lambda3727,3730$ emission lines. The red line represents the Gaussian fit applied to the spectrum.}
    \label{fig:stacked_spectra}
\end{figure*}

The final stacked spectrum, shown in Figure \ref{fig:stacked_spectra}, reveals a mixture of emission lines, primarily originating from the host galaxies, with a clear detection of the [NeV]$\lambda\lambda3346,3426$ line with S/N $\sim 6.7$. The extremities of the spectrum are under-represented due to the redshift distribution, meaning certain features, such as CIII]$\lambda1909$ and [OIII]$\lambda\lambda4959,5007$, are only visible in a subset of sources. In the 3900–4300 $\AA$ range, the continuum is dominated by noise, probably due to imperfect sky subtraction. 

To explore how the redshift might influence emission properties, we measured the EW of the stacked [OII]$\lambda\lambda3727,3730$ line, finding a value of $21.1 \pm 2.1 \AA$. This is in line with the [OII] EW of $18.0 \pm 1.8 \AA$ reported by \citet{2005ApJ...627..721G} for a sample of type 2 Seyfert galaxies, which are associated with lower ionising luminosities. Furthermore, our result is also consistent with the median [OII] EW of $18.8 \pm 1.9 \AA$ found in the SDSS QSO2 sample studied by \citet{2003AJ....126.2125Z} at $0.3 \textless z \textless 0.83$. Thus, the measured [OII] EW in our stacked spectrum aligns well with previous studies.

\subsection{\texttt{LePhare++}: photometric vs spectroscopic redshift}

We estimated the photometric redshifts using the SED fitting code \texttt{LePhare++} to compared them with the ones derived from SDSS spectra. We divide this study into two parts. First, we compared the \texttt{LePhare++} photometric redshifts with the spectroscopic redshift derived from the [OII]$\lambda\lambda3727,3730$ emitters. Then we performed the same analysis with the complete QSO2 candidate sample.

To perform this analysis, we considered the following metrics: the normalised median absolute deviation (NMAD),
\begin{equation}
    \mathrm{NMAD} = 1.48 \, \mathrm{median} \left( \frac{\lvert z_{\mathrm{phot}} - z_{\mathrm{spec}} \rvert}{1+z_{\mathrm{spec}}} \right)
\end{equation}
\noindent to compute the variability between measurements, were $z_{\mathrm{spec}}$ is the spectroscopic redshift from SDSS and z$_{\mathrm{phot}}$ is the photometric redshift derived using \texttt{LePhare++}; the fraction of catastrophic outliers (f$_{\rm{out}}$), as defined in \citet{2010A&A...523A..31H}, where a value is considered to be a catastrophic outlier when
\begin{equation}
    \frac{\lvert z_{\mathrm{phot}} - z_{\mathrm{spec}} \rvert}{1+z_{\mathrm{spec}}} > 0.15;
\end{equation}
\noindent and the bias, where the systematic deviations error are calculated as
\begin{equation}
    \mathrm{bias} = \mathrm{median} \left( \frac{\lvert z_{\mathrm{phot}} - z_{\mathrm{spec}} \rvert}{1+z_{\mathrm{spec}}} \right).
\end{equation}

\begin{table}
\caption{Statistical metrics for the photo-z derived using \texttt{LePhare++}, compared with SDSS spectroscopic redshifts, when considering QSO and galaxies templates.}              
\label{table:metrics_photo_spec_z}      
\centering                                      
\begin{tabular}{c c c c}          
\hline\hline                        
 photo-z  & NMAD & bias & f$_{\rm{out}}$ \\    
\hline                                   
    z$_{\rm{QSO}}$ or z$_{\rm{GAL}}$ & 0.195 & 0.132 & 0.474\\      
    z$_{\rm{GAL}}$ & 0.125 & 0.085 & 0.299 \\
\hline                                             
\end{tabular}
\tablefoot{The photometric redshift output can be selected either by using the best $\chi^2$ value between the QSO templates or the galaxies templates(z$_{\rm{QSO}}$ or z$_{\rm{GAL}}$) or by considering only the best galaxy model (z$_{\rm{GAL}}$). NMAD represents the normalised median absolute deviation, and f$_{\rm{out}}$ indicates the fraction of catastrophic outliers. For all metrics, an ideal value is 0.}
\end{table}

In Table \ref{table:metrics_photo_spec_z}, we present the statistical metrics that compare the z$_{\mathrm{phot}}$ derived with \texttt{LePhare++} to the $z_{\mathrm{spec}}$ from SDSS. We performed two analyses that differ in the selection method for z$_{\mathrm{phot}}$: one based on the best $\chi^2$ value, choosing between z$_{\rm{QSO}}$ (if the QSO template is the best fit) or z$_{\rm{GAL}}$ (if the galaxy template is the best fit), and the other considering only z$_{\rm{GAL}}$. The results show that the estimates of z$_{\mathrm{phot}}$ tend to be systematically lower than the SDSS z$_{\mathrm{spec}}$, with only a few cases where z$_{\mathrm{phot}}$ exceeds the spectroscopic values. This pattern suggests that the combination of optically faint and mid-IR enhanced components introduces degeneracies, denigrating the redshift estimation using SED fitting. In Figure \ref{fig:LePhare_examples}, we show four fitting examples along with their distribution of the probability density function (PDF). 

We should point out that for the best-fit template class, we found that 78\% of sources are best fitted with an AGN template (including composite systems with an AGN component and QSO templates), 28\% of sources are best fitted by a composite template with an AGN component higher than 50\%, and 50\% are best fitted by a QSO template. This result is consistent with the AGN classification presented in C2024, using \texttt{CIGALE}. Due to the degeneracies observed with the \texttt{LePhare++} z$_{\mathrm{phot}}$, we will adopt the spectroscopic redshifts derived by the SDSS pipeline in Section \ref{section:results_cigale}.

\section{\texttt{CIGALE}: physical properties estimation}
\label{section:results_cigale}
Here, we present the results of physical property estimation for our QSO2 candidates using \texttt{CIGALE}. The parameter grids for the control sample and the QSO2 candidates are shown in Tables \ref{tab:SED_fitting_params_gal} and \ref{tab:SED_fitting_params_QSO2}, respectively. We took into account the parameters provided in \citet{2020MNRAS.491..740Y} and modified them according to our needs. The main reason for using two different grids for the control sample and the QSO2 candidates is that, when the same grid was applied to both, we observed that the derived SFRs were hitting a plateau at the upper limit. We found that for our control sample, a more recent starburst event was required to explain the bright UV emission. Therefore, a more recent SFH was required to successfully fit the control sample and derive its physical properties, compared to the QSO2 candidates. We adjusted the stellar age components for SFH, allowing the inference to converge and provide a more reliable estimation.

\subsection{AGN classification with $\mathrm{frac_{AGN}}$}

Since \texttt{CIGALE} uses a multi-component models to fit the SEDs of galaxies and a Bayesian approach to refine the output properties, it allows to derive well-constrained physical properties and uncertainties, under the assumption that the model we fit are representative of the sample under consideration. We take advantage of the implementation of the \texttt{SKIRTOR} model to infer the best-fitting parameters related to the AGN component. The AGN model will be of high importance for fitting the enhanced IR emission observed in our candidate sample. In Fig. \ref{fig:SED_example}, we show four examples of the best-fit model using \texttt{CIGALE}.

As previously observed in C2024, the \texttt{SKIRTOR} model consistently estimates a high frac$_{\mathrm{AGN}}$. In this study, we will also consider a source to have a substantial AGN contribution when frac$_{\mathrm{AGN}} > 0.1$, following the criteria established by \citet{2018ApJ...854...62L} and \citet{2022MNRAS.509.4940T}. This criterion is satisfied by all members of our candidate sample, aligning with their AGN classification. Furthermore, we applied the \texttt{SKIRTOR} model to the control sample to study its ability to recover the frac$_{\mathrm{AGN}}$ in sources without a considerable AGN contribution. We found that all galaxies within the control sample exhibited a frac$_{\mathrm{AGN}} \leq 0.1$, consistent with our expectation that these are primarily star-forming galaxies without a significant AGN contribution.

\begin{figure}
    \centering
    \includegraphics[width=\linewidth]{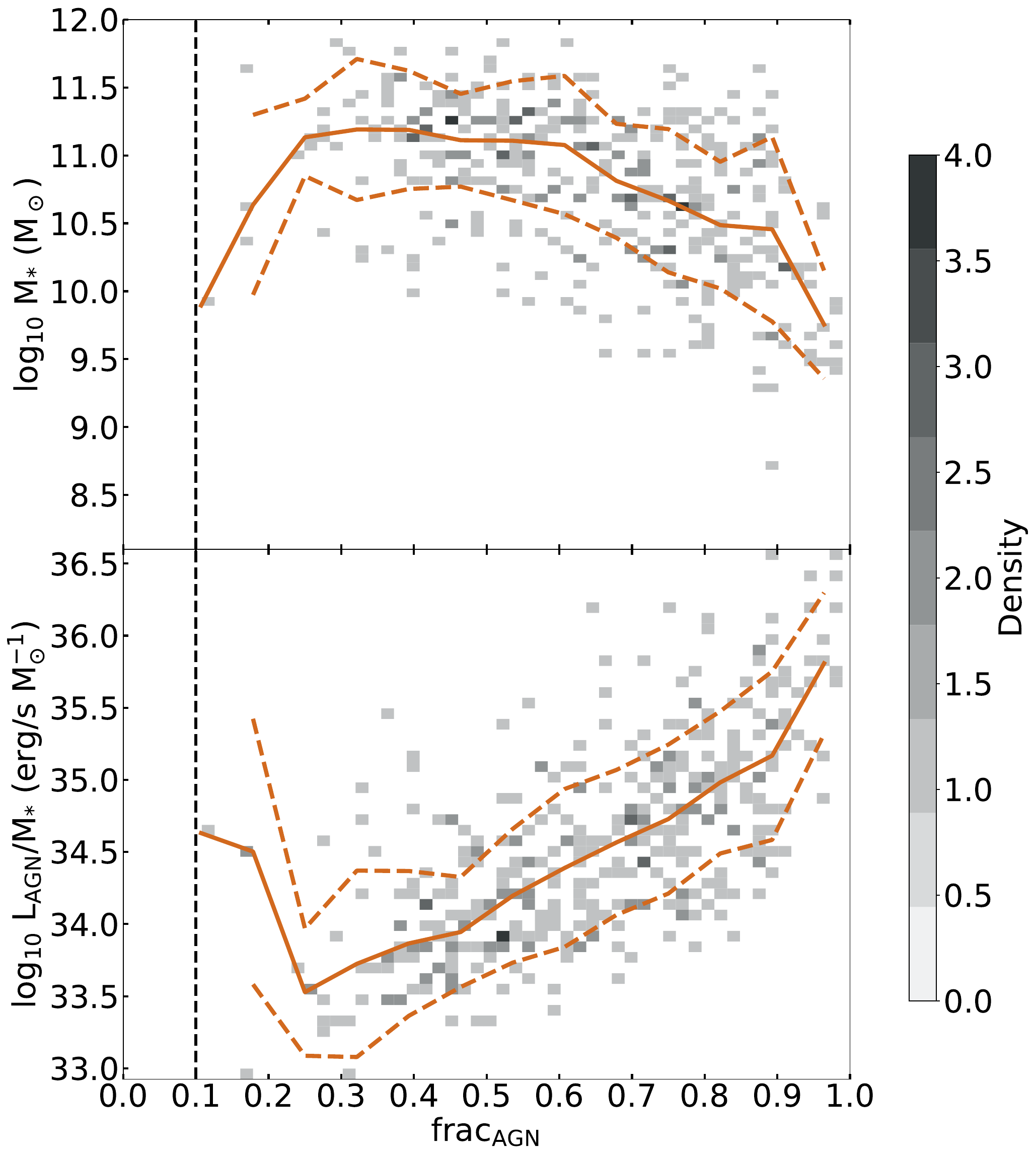}
    \caption{2D histogram of the stellar mass, M$_{\ast}$, (top) and specific AGN luminosity, L$_{\mathrm{AGN}}$/M${*}$, defined in \citet{2022MNRAS.509.4940T}, (bottom) as a function of frac$_{\mathrm{AGN}}$ for the QSO2 sample. The colour gradient indicates the density of sources for each bin. The median and standard deviation are represented by the solid and dashed orange lines, respectively. The vertical black line indicates the frac$_{\mathrm{AGN}} > 0.1$ threshold used in \citet{2018ApJ...854...62L}, to characterise sources with significant AGN contribution.}
    \label{fig:SED_sLAGN}
\end{figure}

The specific AGN luminosity is defined as the ratio between the AGN luminosity (L$_{\mathrm{AGN}}$) and the stellar mass (M$_*$): L$_{\mathrm{AGN}}$/M$_*$ \citep{2016MNRAS.460..902B,2022MNRAS.509.4940T}. In Fig. \ref{fig:SED_sLAGN}, we show a 2D histogram of M$_{\ast}$ and L$_{\mathrm{AGN}}$/M$_{\ast}$, as a function of frac$_{\mathrm{AGN}}$ of our sample. For analysis purposes, we plot the median and standard deviation of each physical property, as solid and dashed orange lines, to guide the reader. While studies focusing on the stellar mass (M$_{\ast}$) distribution of QSO2s are limited, \citet{2014MNRAS.438.1839B} estimated the M$_{\ast}$ of the QSO2 SDSS J002531-104022 to be 4-17$\times 10^{10}$ M$_{\odot}$, which is within our derived M$_{\ast}$ distribution interval.

When looking at how frac${\mathrm{AGN}}$ varies with M$_{\ast}$, we see that M$_{\ast}$ decreases as frac${\mathrm{AGN}}$ increases. This indicates a significant anti-correlation: less massive galaxies tend to have a higher fractional AGN contribution. This trend is supported by a Spearman's rank correlation coefficient of -0.488 (p < 0.001), showing a moderate negative relationship between the two variables. 

Considering that only nine photometric points are being used in this study, due to the lack of observations in other wavelength regimes (including FIR), one can argue that \texttt{CIGALE} might be assigning most of the total luminosity to the AGN component, resulting in an underestimation of the stellar mass. We investigated the relationship between AGN activity and stellar mass through two approaches. First, for sources with high AGN contribution (frac$_{\mathrm{AGN}} > 0.5$), the Pearson correlation coefficient between AGN luminosity and stellar mass was -0.02, indicating no significant linear relationship. Second, expanding to the full sample (all frac$_{\mathrm{AGN}}$ values), the coefficient was -0.17, showing a very weak negative correlation. A visual inspection of the frac$_{\mathrm{AGN}}$ vs AGN luminosity plot, colour-coded by stellar mass, focused on AGN-dominated sources (frac$_{\mathrm{AGN}} > 0.5$), revealed no clear trends or evidence that AGN luminosity affects the estimation of stellar mass.

A possible justification is the role of AGN feedback, through outflows or radiation pressure that can heat and expel gas from the galaxy, effectively suppressing star formation \citep{2024Galax..12...17H}. This effect is more significant in galaxies with a higher frac${\mathrm{AGN}}$, where the AGN energy output plays a dominant role \cite[e.g.][]{2015ApJ...799...82C}. The observed anti-correlation remains consistent even when focusing on a subsample of galaxies within a narrow range of M$_{\ast}$.  

On the other hand, we identify a positive correlation in L$_{\mathrm{AGN}}$/M$_{\ast}$ with frac$_{\mathrm{AGN}}$. \citet{2022MNRAS.509.4940T} also identifies such trend for sources with frac$_{\mathrm{AGN}}$ > 0.8. Since our sample is much smaller and represents a specific subtype of AGNs, we do not expect the same trend in \citet{2022MNRAS.509.4940T} to be present. An important caveat to remember is that only optical and IR data, from 0.3-22 $\mu$m, is being used in this study. Therefore, because of optical obscuration observed in our sample, we expect the IR part of the SED to dominate, which increases L$_{\mathrm{AGN}}$ with frac$_{\mathrm{AGN}}$.

A significant caveat of our study is the absence of far-infrared (FIR) data, which would support the derivation of accurate SFRs, as noted by \citet{2015A&A...576A..10C}. Consequently, the estimation of AGN luminosity would also be improved by incorporating FIR and sub-millimetre data. We conducted a 4\arcsec radius search of the updated Herschel/PACS Point Source Catalogue \citep{2024A&A...688A.203M} but did not find any counterparts. Since both C2024 and this current work focus on the analysis of QSO2 candidates selected from a large survey focused on the Northern Hemisphere, and therefore covering a significant area of the sky, we are directly impacted by the limited availability of all-sky multi-wavelength data. Thus, our analysis provides a preliminary exploration of the physical properties of our candidate sample, forming the basis for future observational time proposals to further characterise our QSO2 candidates, in the redshift desert.

\subsection{SFR-M$_{\ast}$ diagram}
\label{section:SFR_M*}

The intrinsic connection between SFR and M$_{\ast}$ has been extensively discussed in the literature as a key relationship to study and understand galaxy evolution \citep[e.g.][and references therein]{2000ApJ...536L..77B,2004MNRAS.351.1151B,2007ApJ...660L..43N,2018A&A...609A..82B}. \citet{2015A&A...575A..74S} parameterise the SFR of main sequence (MS) galaxies as a function of redshift and stellar mass:

\begin{equation}
    \log_{10}(\text{SFR}_{\text{MS}}[M_{\odot}/\text{yr}]) = m - m_0 + a_0 r - a_1 \left[\max(0, m - m_1 - a_2 r)\right]^2
\end{equation}

with $m_0$ = 0.5 $\pm$ 0.07, $a_0$ = 1.5 $\pm$ 0.15, $a_1$ = 0.3 $\pm$ 0.08, $m_1$ = 0.36 $\pm$ 0.3, $a_2$ = 2.5 $\pm$ 0.6, \(r = \log_{10}(1+z)\) where \(z\) is the redshift, and \(m = \log_{10}(\rm{M_*/10^9 M_{\odot}})\). The parameters are physically motivated to preserve the slope increase with redshift and the "bending" effect, verified at low stellar mass and high redshift \citep[see][for more details]{2015A&A...575A..74S}.

\begin{figure}
    \centering
    \includegraphics[width=\linewidth]{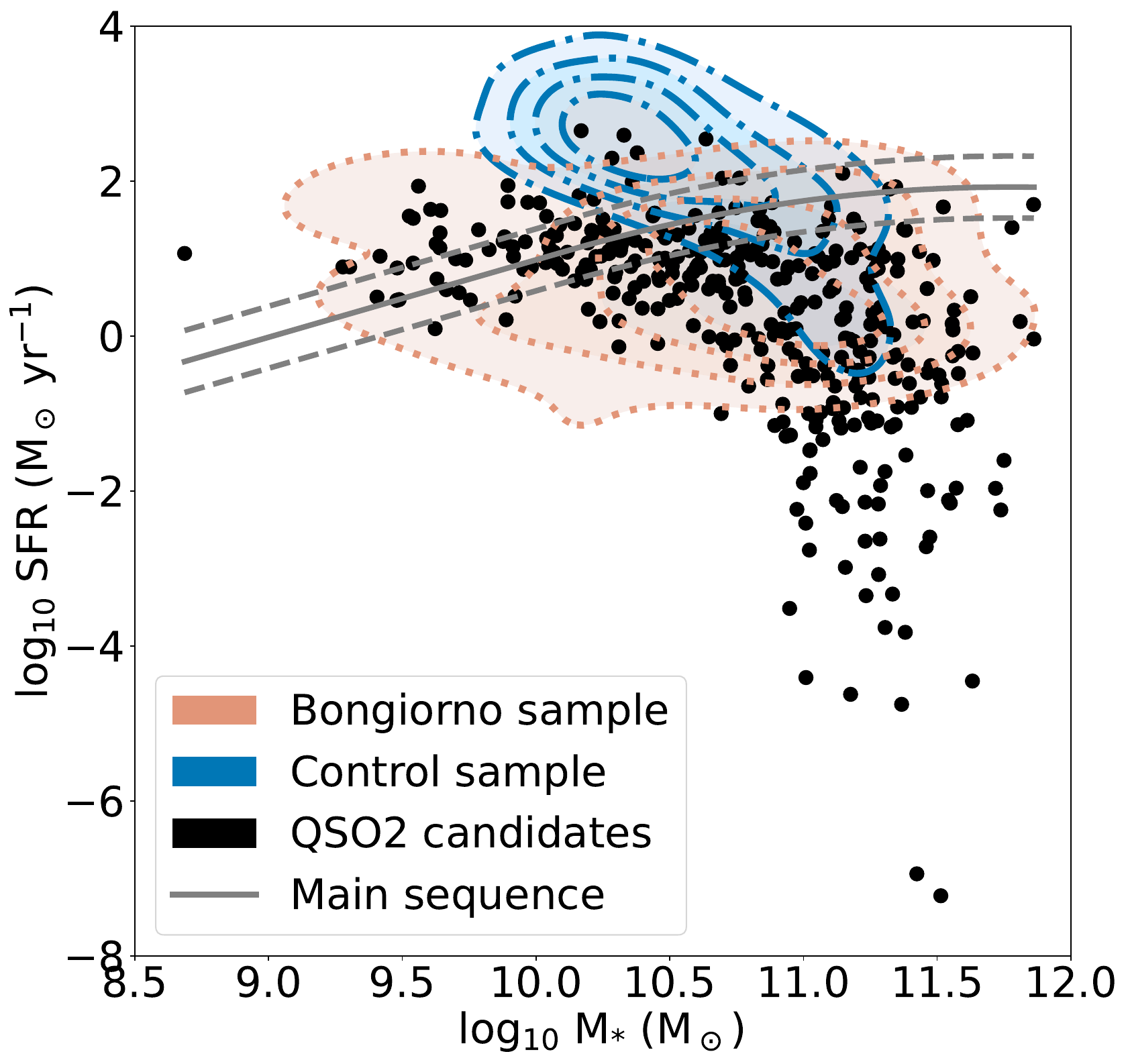}
    \caption{The star formation rate (SFR) as a function of stellar mass (M$_*$) for the QSO2 candidates (black dots) is plotted on a logarithmic scale. The star formation main sequence (MS) described by \citet{2015A&A...575A..74S} at the mean SDSS spectroscopic redshift of our candidates (\(z = 1.1\)) is shown as a solid line, with dashed lines indicating a 0.4 dex deviation. The contours display the M$_*$-SFR distribution for X-ray and optically selected AGN2 from the zCOSMOS survey by \citet{2012MNRAS.427.3103B} with redshifts \(1 \leq z \leq 2\) (orange) and for the control sample of SDSS galaxies (blue).}
    \label{fig:SED_Mstellar_SFR}
\end{figure}

\begin{figure}
    \centering
    \includegraphics[width=\linewidth]{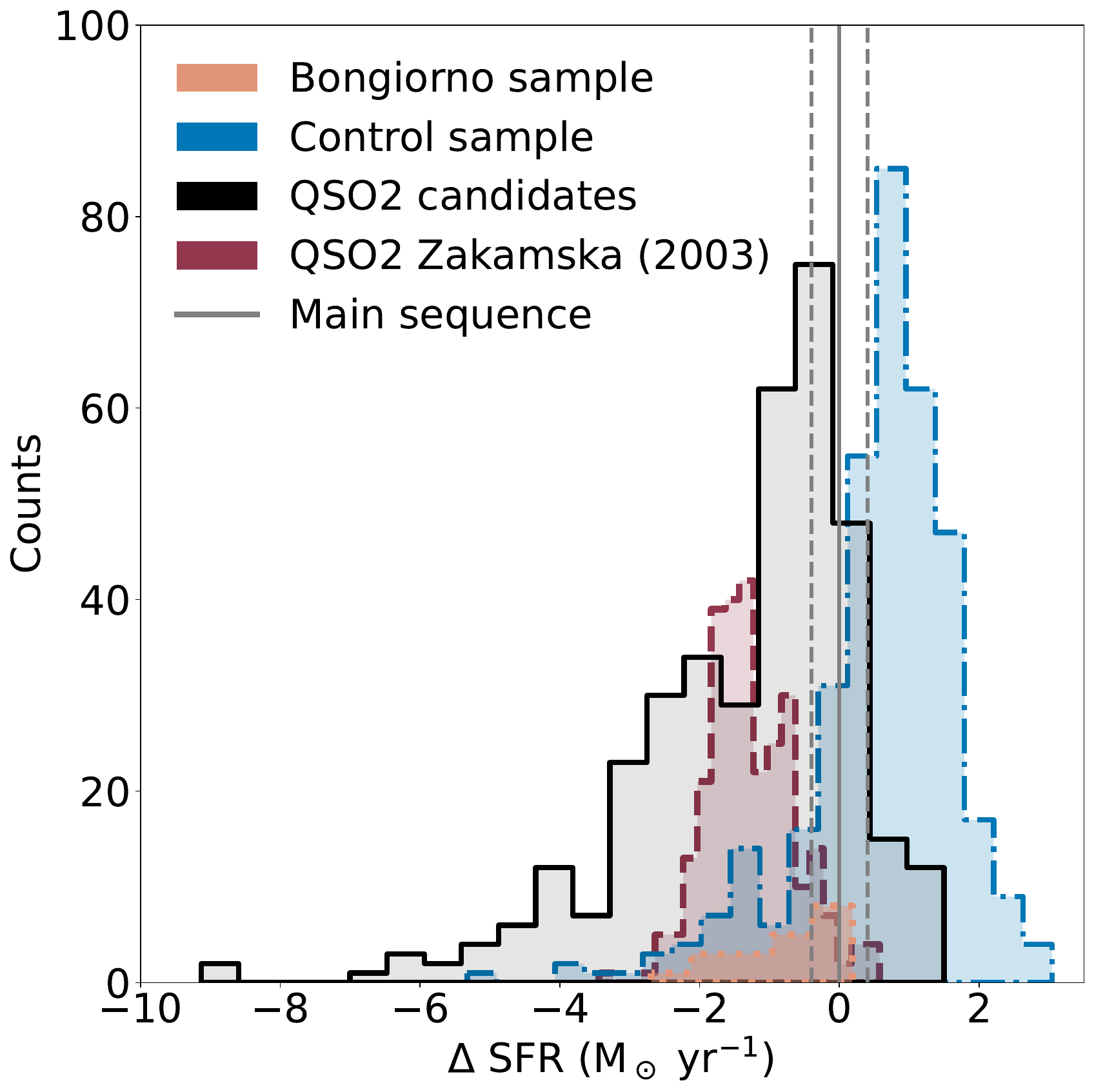}
    \caption{Difference between the estimated SFR and the main sequence SFR for the QSO2 candidates, the control sample, the Bongiorno sample \citep{2012MNRAS.427.3103B}, and the Zakamska sample \citep{2003AJ....126.2125Z}. The solid grey line represents the main sequence of star formation found by \citet{2015A&A...575A..74S} at \(z = 1.1\), and the dashed lines give a scatter of $\pm$ 0.4 dex.}
    \label{fig:SED_delta_SFR}
\end{figure}

In Figure \ref{fig:SED_Mstellar_SFR}, we plot the SFR-M$_{\ast}$ diagram for the QSO2 candidates and SDSS galaxies. The star formation main sequence (MS) described by \citet{2015A&A...575A..74S}, at the mean SDSS spectroscopic redshift of our candidates (\(z = 1.1\)), is plotted as a solid line. In the contours, we show the SFR-M$_{\ast}$ distribution for a sample of 87 X-ray and optically selected AGN2 from the zCOSMOS survey by \citet{2012MNRAS.427.3103B} with redshift \(1 \leq z \leq 2\). This analysis allows us to understand and compare the distribution of our candidates with confirmed AGN2. We observe a similar distribution in the SFR-M$_{\ast}$ diagram, with some candidates having higher M$_{\ast}$ and SFR. For the SDSS control sample, we observe, on average, a higher SFR estimation within similar M$_{\ast}$.

In \citet{2003AJ....126.2125Z}, a sample of 291 QSO2, at redshift between 0.3 and 0.83, using SDSS spectroscopic data. These sources were identified by their narrow emission lines (FWHM \textless 2000 km/s), and their high EWs and high-ionisation lines ratios (e.g. [OIII]$\lambda5008$/H$\beta\lambda4861$). We used the available [OII] luminosities\footnote{\url{https://cdsarc.cds.unistra.fr/viz-bin/cat/J/AJ/127/2002}} to compute their SFR using the \citet{2004AJ....127.2002K} relation:
\begin{equation}
\rm{SFR}([\rm{OII}])(M_{\odot} yr^{-1}) = (6.58 \pm 1.65) \times 10^{-42} \rm{L([OII])} \text{ ergs } s^{-1}
\end{equation}

In Fig. \ref{fig:SED_delta_SFR}, we plot the difference between the \texttt{CIGALE} estimated SFR and the MS SFR for the QSO2 candidates, the control sample, the Bongiorno sample \citep{2012MNRAS.427.3103B}, and the Zakamska sample \citep{2003AJ....126.2125Z}. It is clear that our candidate sample has a broader distribution of SFR, encompassing a mixture of starburst to quiescent galaxies. Regarding the higher SFR seen for the control sample, selection bias can play a role here, as for star-forming galaxies to be detected at these redshifts, the [OII] $\lambda\lambda 3727, 3730$ lines must be bright, which in itself requires higher SFR. Finally, we observe that our sample of QSO2 candidates is spread across the SFR-M$_{\ast}$ diagram, from the starburst to the quiescent regime.


\section{\texttt{CIGALE}: Multi-wavelength analysis}
\label{section:multiwavelength}
The SED fitting results presented above are based on optical to MIR photometry, but it is well-known that AGNs have clear signatures outside these wavelength ranges. For a subset of our candidates we can check our SED fitting results by incorporating radio and/or X-ray data, and in the following we will discuss the resulting impact on the inferred physical parameters.

\subsection{Including LoTSS radio photometry}
We tested the inclusion of LoTSS radio photometry \citep{2022A&A...659A...1S} alongside SDSS and WISE data to assess its impact on the fit in \texttt{CIGALE}. A 5\arcsec radius search of our coordinates against the LoTSS Data Release 2\footnote{\url{https://lofar-surveys.org/public/DR2/catalogues/LoTSS_DR2_v110_masked.srl.fits}} identified matches with 16 sources. Our primary objective was to evaluate how the inclusion of radio data influences the estimation of AGN physical properties, particularly by altering the IR fit and its effect on the inferred galaxy properties.

\begin{figure}
    \centering
    \includegraphics[width=\linewidth]{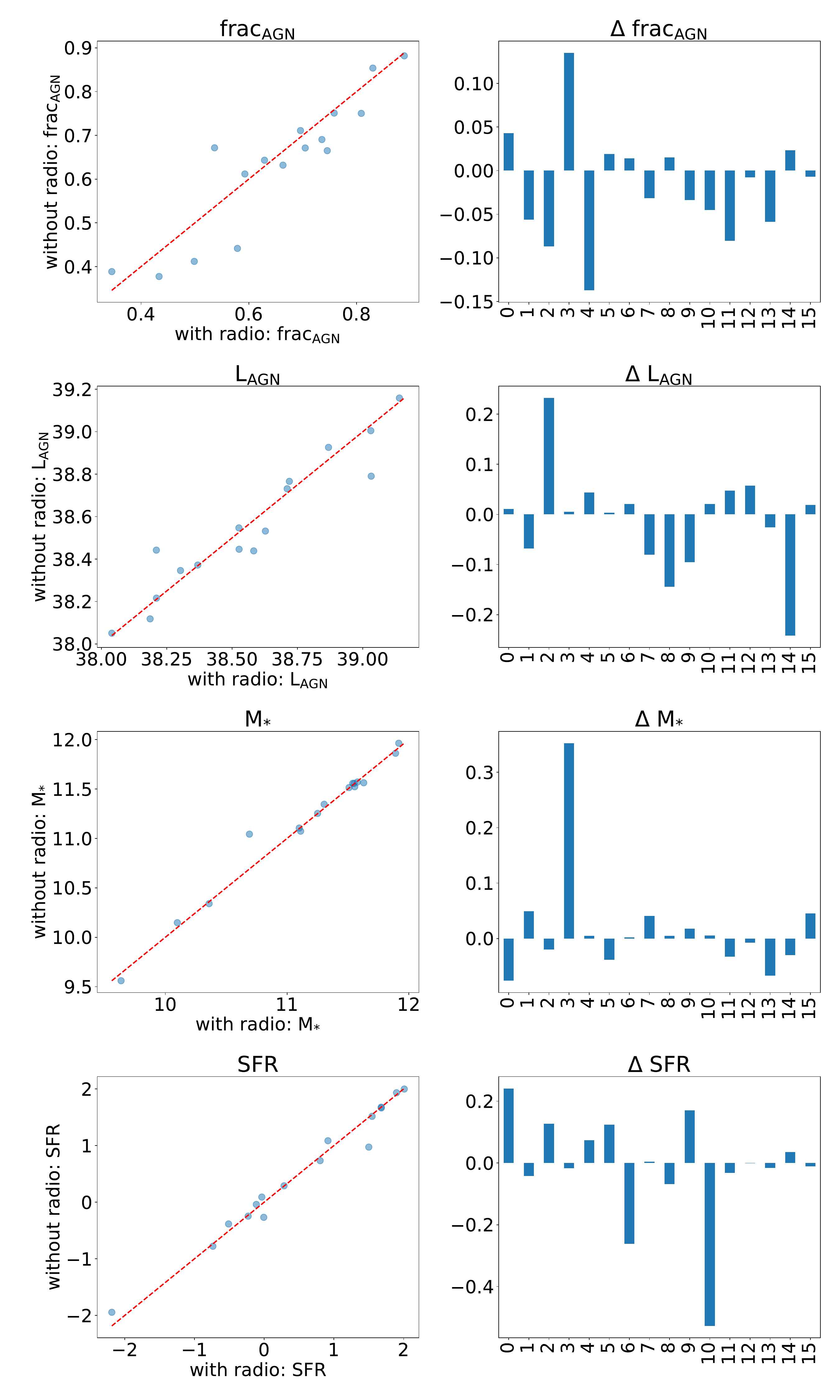}
    \caption{Comparison between physical properties estimated with and without radio LoTSS photometry, using \texttt{CIGALE}. The left column shows scatter plots comparing various galaxy properties such as AGN fraction, AGN luminosity, stellar mass, and star formation rate between galaxies with radio LoTSS photometry and without. All values are provided in logarithmic scale, with the exception of the  frac$_{\mathrm{AGN}}$. Each scatter plot includes a 1:1 reference line, as a dashed red line. The right column presents bar plots of the differences in the same properties, for each source (numbered in the x axis): SDSS+WISE - SDSS+WISE+LoTSS.}
    \label{fig:compare_radio}
\end{figure}

In Fig. \ref{fig:compare_radio}, we show the scatter plot with the physical properties with and without radio photometry. Additionally, we show the difference between the physical properties estimated with radio photometry subtracted by those estimated without. We observed that the properties of the host galaxies, M$_{\ast}$ and SFR, show only slight variations, which do not have a profound impact on the overall study. When including the radio data, the frac$_{\mathrm{AGN}}$ and L$_{\mathrm{AGN}}$ inferred are somewhat higher than if radio data are excluded. In particular, we see a tendency to estimate a smaller frac$_{\mathrm{AGN}}$. The root cause of this difference is that when including radio data, there is tension between the non-thermal radio model fit and the mid-IR part of the SED fit, leading to the latter being less well constrained. This signals a potential inconsistency in the MIR to radio modelling and to make progress a more in-depth study of this is needed with potentially updates to the model, this is however outside the bounds of this work.

\subsection{Including XMM-Newton X-ray photometry}
Taking advantage of the flexibility provided by \texttt{CIGALE}, we conducted a similar comparison to evaluate the impact of incorporating X-ray photometry into the SED fitting process. A 2\arcsec radius search for our QSO2 candidates identified matches with five sources. We then integrated the 2–12 keV XMM-Newton photometry \citep{2012ApJ...756...27L} into \texttt{CIGALE} for analysis.

\begin{figure}
    \centering
    \includegraphics[width=1\linewidth]{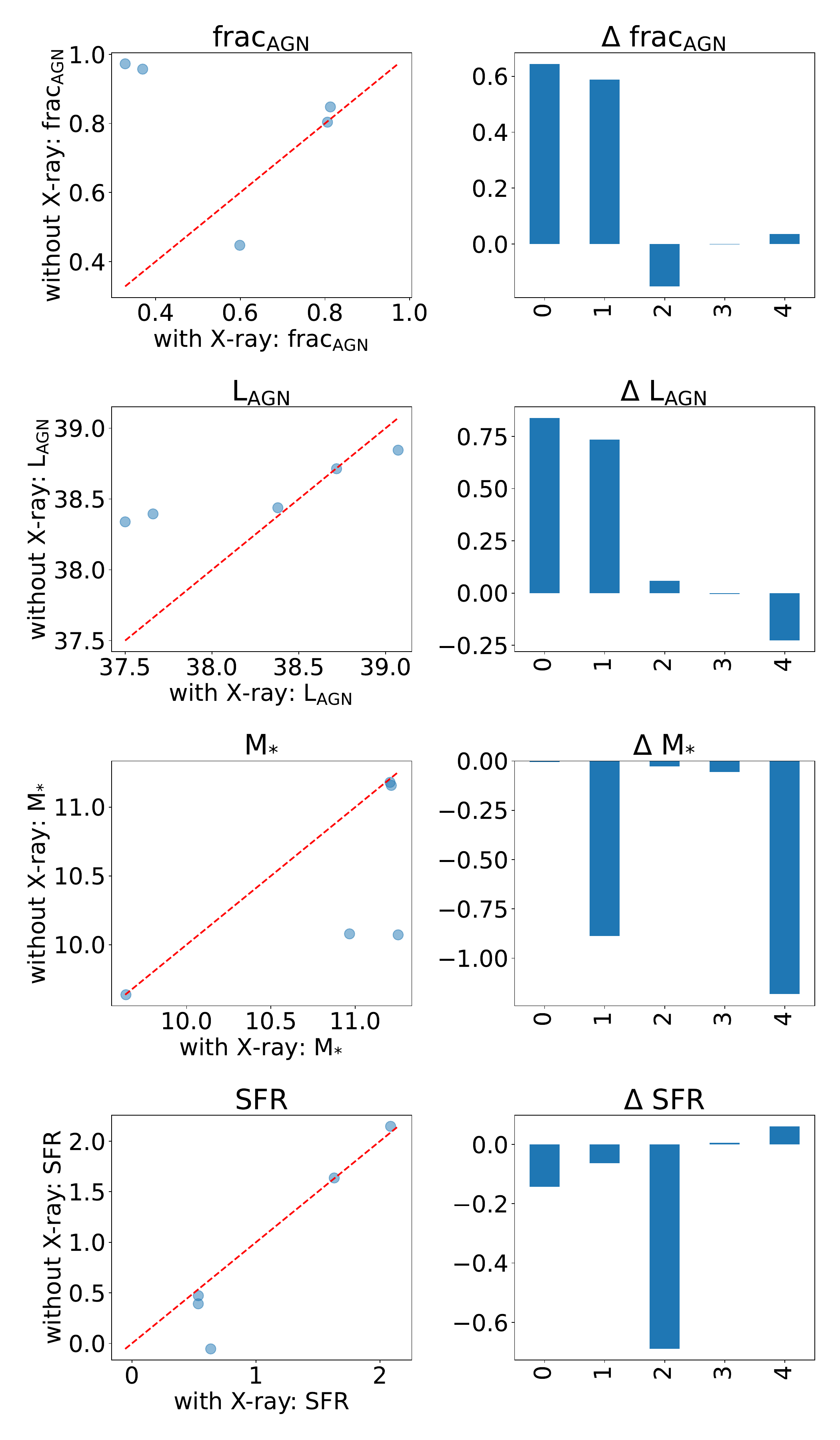}
    \caption{Comparison between physical properties estimated with and without X-ray XMM-Newton photometry, using \texttt{CIGALE}. The left column shows scatter plots comparing various galaxy properties such as AGN fraction, AGN luminosity, stellar mass, and star formation rate between galaxies with X-ray XMM-Newton photometry and without. All values are provided in logarithmic scale, with the exception of the  frac$_{\mathrm{AGN}}$. Each scatter plot includes a 1:1 reference line, as a dashed red line. The right column presents bar plots of the differences in the same properties, for each source (numbered in the x axis): SDSS+WISE - SDSS+WISE+XMM-Newton.}
    \label{fig:compare_xray}
\end{figure}

The comparison study is presented in Fig. \ref{fig:compare_xray}, similar to Fig. \ref{fig:compare_radio}, where the physical properties of five X-ray emitters are presented. In this case, we see that the introduction of X-ray data changes the estimated properties for both the AGN and the host galaxy. The parameter less affected is the SFR, which increases for sources with a lower estimated SFR, using SDSS+WISE. For two of the five sources, the stellar mass estimates when including X-ray data increase by nearly 1 dex, indicating a significant underestimate of the stellar mass when X-rays are included. 

The physical properties of AGN have the highest variance in this study. For two sources, the L$_{\mathrm{AGN}}$ is overestimated, according to the one estimated with the addition of X-ray photometry. The most affected property is frac$_{\mathrm{AGN}}$, with two sources having a difference of $\sim 0.6$, resulting in frac$_{\mathrm{AGN}} \sim 0.4$. Although the AGN contribution remains significant, the observed difference in AGN physical properties arises from the reduced fit to the IR part of the SED when the X-ray model is included. When X-ray photometry is included, the fit places less weight on the mid-IR, resulting in an underestimate of the mid-IR emission. This leads to a decrease in frac${\mathrm{AGN}}$ while compensating for the increase in stellar mass.


\section{Separating QSO2 candidates using SFR}
\label{section:SFR_separation}

The ratio between SFR and SFR$_{MS}$ allows us to create different regions that separate galaxies into starbursts, star formation, and quiescent \citep [e.g.][]{2015ApJ...806..187A, 2019MNRAS.484.4360A, 2022A&A...659A.129V}. In our sample, the galaxies are divided into four regimes: starburst, log(SFR/SFR$_{MS}$) > 0.4 dex; star-forming, log(SFR/SFR$_{MS}$) $\pm$ 0.4 dex; sub-MS, log(SFR/SFR$_{MS}$) between -0.4 and -1.3; and quiescent, log(SFR/SFR$_{MS}$) < -1.3 \citep[e.g.][]{2019MNRAS.484.4360A,2022A&A...659A.129V}. 

\begin{figure}
\centering
\includegraphics[width=\linewidth]{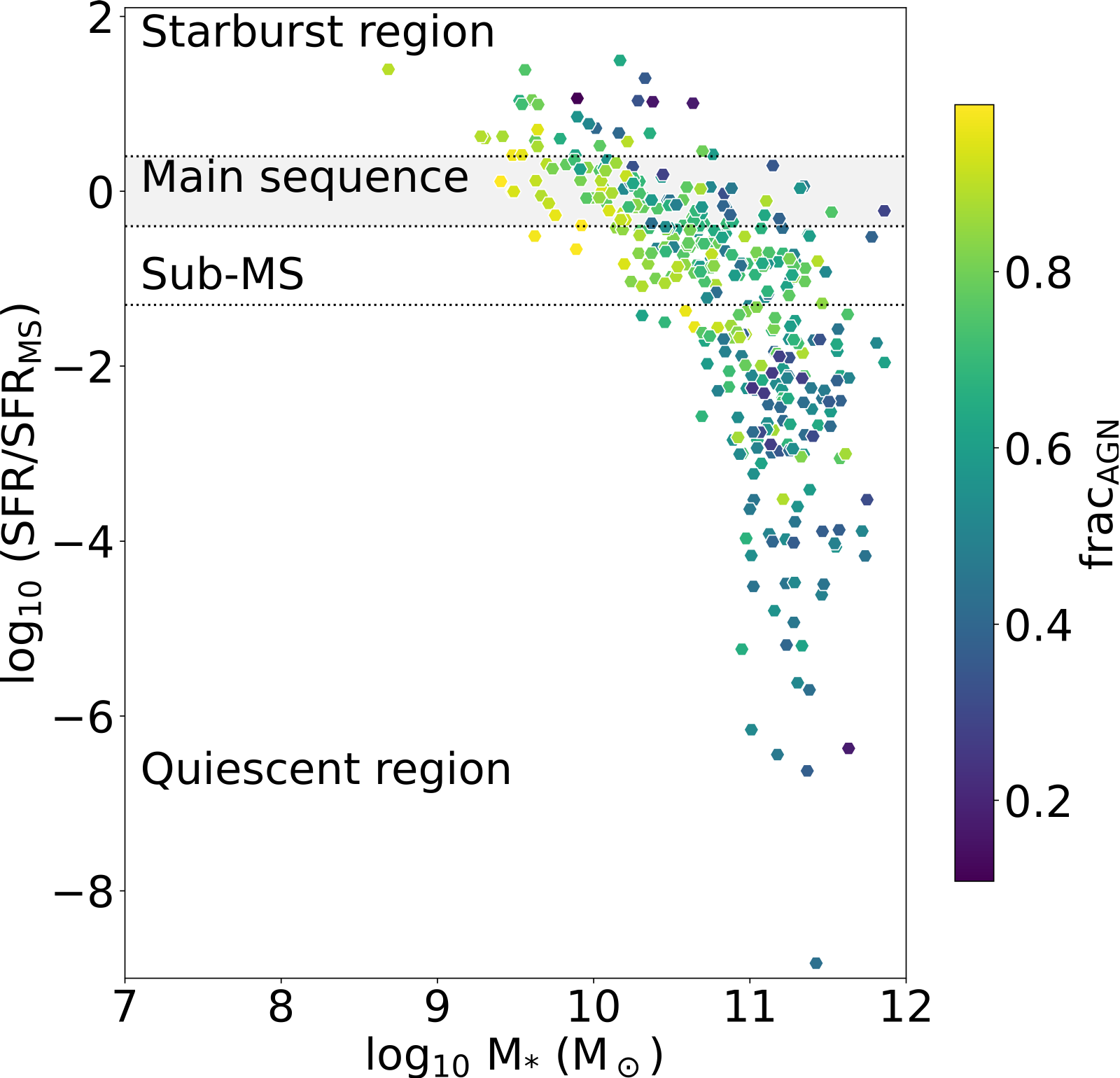}
\caption{Measurements of the logarithm of the ratio SFR/SFR$_{MS}$ as a function of the logarithm of stellar , M$_{\ast}$. The SFRs relative to the main sequence of star formation for galaxies were computed assuming the average redshift from our QSO2 candidate sample, z=1.1 and using the relation by \citet{2015A&A...575A..74S}. The dashed line separates the main sequence and the starburst region at 0.4 dex, as defined by \citet{2019MNRAS.484.4360A}. The blue-shaded area defines the main sequence region. Each point is colour coded based on the estimated AGN fraction of CIGALE, frac$_{\mathrm{AGN}}$.}
\label{fig:SED_Mstellar_SFR_SFRmain}
\end{figure}

To understand how the quantity SFR/SFR$_{MS}$ relates to the M$_{\ast}$, we show the results for our QSO2 candidates in Figure \ref{fig:SED_Mstellar_SFR_SFRmain}. The sources are colour-coded according to the estimated frac$_{\mathrm{AGN}}$. In summary, we obtain 31 sources in the starburst region ($\sim 8.49 \%$); 88 sources in the star-forming region ($\sim 24.11 \%$); 98 sources in the sub-MS region ($\sim 26.85 \%$); and 148 sources in the quiescent region ($\sim 40.55 \%$). The candidates in the quiescent region are among the most massive candidates with M$_{\ast}$ within the range 10$^{10.5-12}$ M$_{\odot}$, while the starburst and main sequence regions show a wider M$_{\ast}$ range. Similarly to the results of \citet{2022A&A...659A.129V}, we do verify a preference for massive host galaxies (M$_{\ast}$ $\ge$ 10$^{10}$ M$_{\odot}$) to have a lower SFR. 

Due to the proximity of the redshift of our sample to the peak of star formation (z $\sim 2$, \citealt{2014ARA&A..52..415M}), the diversity of the estimated SFR is noteworthy. For example, we observe sources in the starburst region with M$_{\ast} \sim 10^{9-11}$ and a mean frac$_{\mathrm{AGN}} \sim 0.7$. Starburst galaxies have been reported as a direct consequence of major merging processes \citep[e.g.,][]{1996ARA&A..34..749S,2006asup.book..285L}, which can trigger AGN activity \citep[e.g.,][]{1988ApJ...325...74S, 2005Natur.433..604D, 2006ApJS..163....1H}, explaining the high SFR and dust content. Furthermore, we obtain a high percentage of sources in the quiescent region, with a median frac$_{\mathrm{AGN}} \sim 0.6$. These can be sources with obscured star-formation due to the dusty nature of galaxy \citep[e.g.][]{1999MNRAS.309..715B,2007ApJ...670..173D,2008ApJ...687..835A}, therefore being a direct consequence of a merger-driven system, or it can be a consequence of AGN feedback which may be redistribution of the gas content of the galaxy, diminishing the ability of the galaxy to produce new stars \citep[e.g.][]{2012AdAst2012E..16T,2018A&A...620A.193V,2024Galax..12...17H,2025ApJ...981...25B}. Another explanation can be a 'cocooned´ AGN with Compton thick obscuration (N$_{\rm{H}} \geq 10^{24}$ cm$^{-2}$), whose optical emission is absorbed and enshrouded by gas and dust \cite[also known as optically quiescent quasars (OQQs),][]{2021MNRAS.503L..80G,2022ApJ...934L..34G}.

\begin{figure}
    \centering
    \includegraphics[width=\linewidth]{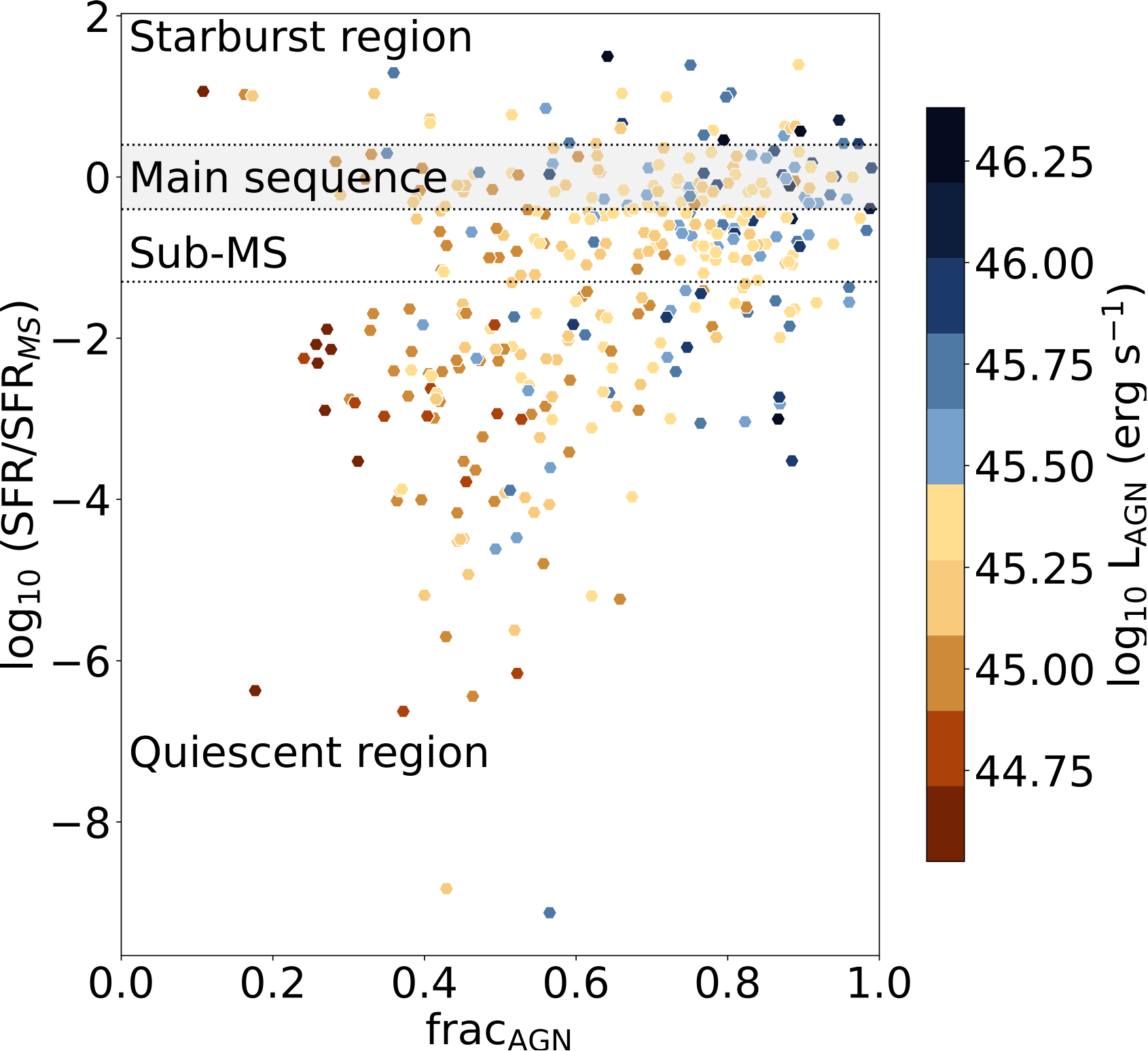}
    \caption{Similar plot to Figure \ref{fig:SED_Mstellar_SFR_SFRmain}, but with SFR/SFR$_{MS}$ as a function of frac$_{\mathrm{AGN}}$. Each point is colour-coded based on the \texttt{CIGALE} estimated log$_{10}$ L$_{\rm{AGN}}$.}
    \label{fig:SED_fracAGN_SFR_SFRmain}
\end{figure}

To further study the distribution of the frac\(_{\mathrm{AGN}}\) parameter, Figure \ref{fig:SED_fracAGN_SFR_SFRmain} shows the distribution of SFR/SFR\(_{MS}\) as a function of frac\(_{\mathrm{AGN}}\) with the total luminosity of AGN, L\(_{\mathrm{AGN}}\). Galaxies within the star-forming MS present a wide range of frac\(_{\mathrm{AGN}}\), similar to the fraction of galaxies in the starburst region. We observe an increase in L\(_{\mathrm{AGN}}\) with frac\(_{\mathrm{AGN}}\), expected due to the contribution of the IR emission, with AGN-like luminosities between 10\(^{44-47}\) erg s\(^{-1}\). In the MS region, galaxies are luminous and have a high frac\(_{\mathrm{AGN}}\). 

The detection of emission from star formation, while presenting apparent dust emission in the IR, matches the description of AGN2 found as [NeV]\(\lambda3426\) emitters \citep[e.g.][]{2018A&A...620A.193V}, known as composite systems with a combined contribution of star formation and AGN to the electromagnetic spectrum, also presented in C2024. These results are also consistent with \citet{2021AJ....162...65H}, where AGN2 are typically found on or near the star-forming MS, and \citet{2022MNRAS.514.2936W}, where cosmological simulations predict at \(0 \leq z \leq 2\) that AGNs are preferentially found in star-forming galaxies.

\section{Comparison with AGN2 samples from the literature}
\label{section:comparison_literature}
In \citet{2022A&A...659A.129V}, optically selected AGN2 with redshifts between 0.5 and 0.9 from the VIPERS and VVDS surveys was studied to better understand the connection between AGN activity and the physical properties of the host galaxies. Similarly to this work, in \citet{2022A&A...659A.129V} the \texttt{CIGALE} code was used to derive the stellar masses, while the SFR was computed using the [OII]$\lambda$3726+3728 doublet line. We compared our results to understand how the physical properties of our candidates are distributed. For stellar masses higher than 10$^{10}$ M$_{\odot}$, the majority of our sources are also below the MS.

The distribution of the Bongiorno sample and \citet{2022A&A...659A.129V} on the SFR-M$_{\ast}$ diagram is very similar, with the majority of AGN2 clustered below the MS region and log$_{10}$SFR $\leq$ 1.5. Both studies based their classification of AGN on standard diagnostic diagrams [OIII]/H$\beta$ versus [OII]/H$\beta$ \citep[see][and references therein]{1997MNRAS.289..419R, 2010A&A...509A..53L}. Therefore, their selection bias will provide sources with lower SFR as the majority of star-forming galaxies will be excluded from the sample \footnote{In \citet{2022ApJ...925...74J}, the authors discuss a similar effect, where AGN selection methods for X-ray AGNs and IR AGNs show different physical properties for the AGN host galaxies}. Nevertheless, from Fig. \ref{fig:SED_Mstellar_SFR}, our selection provides us with a selection of sources with lower SFR than the aforementioned samples.

In this study, we constrain the photometry of our sample to preferentially select faint optical sources that align with the QSO2 sample from \citet{2013MNRAS.435.3306A}, as detailed in Section \ref{section:Data_qso2}. This selection criterion will directly affect the estimated optical contribution to the SFR, enabling the infrared (IR) contribution to be treated as a free parameter, that is, no constraints are applied to the IR data from WISE. Consequently, the results presented in Fig. \ref{fig:SED_Mstellar_SFR} become particularly pertinent, as the optical contribution to the SFR estimation will be relatively minor compared to the IR contribution. Through our methodology, we identify QSO2 candidates across a broader range of SFRs, thus presenting a complex sample that potentially reflects different evolutionary mechanisms driving the observed properties.

\section{Comparison with simulations: \texttt{SPRITZ}}
\label{section:SPRITZ}
We performed a comparison with the semi-empirical simulated catalogue Spectro-Photometric Realisations of Infrared-selected Targets\footnote{\url{http://spritz.oas.inaf.it/catalogs.html}} \citep[\texttt{SPRITZ};][master catalogue v1.131]{2021PASA...38...64B,2021A&A...651A..52B,2022A&A...666A.193B}. In this catalogue, luminosity functions and galaxy populations are combined with SED simulated galaxies to reproduce physical properties, emission features, and expected fluxes using SED fitting for all simulated sources \citep[see][ and references therein]{2021A&A...651A..52B}. The galaxy populations are classified in \texttt{SPRITZ} based on the observations results presented in \citet{2013MNRAS.432...23G}:
\begin{itemize}
    \item Non-AGN dominated: star-forming (SF) galaxies, with no clear AGN activity, which includes spiral and dwarf galaxies; starburst (SB) galaxies; and passive elliptical galaxies with little or no star-formation;
    \item AGNs: unobscured AGN (AGN1); and luminous galaxies with an AGN component dominating over the star-forming activity, with UV-optical obscuration due to dust (AGN2);
    \item Composite systems: SB-AGN typically have highly obscured AGN or Compton-thick (N$_{\rm{H}}\geq 10^{23.5}cm^{-2}$) with a non-negligible component of star formation, and enhanced mid-IR emission; SF-AGN with properties similar to a spiral galaxy while having a low-luminosity or obscured AGN, with a "flattening" in the 3-10$\mu$m range. Furthermore, SF-AGN(SB) and SF-AGN(spiral) are divided based on their far-IR/near-IR colours and SED similarities;
\end{itemize}

 For this work, we are particularly interested in the following subgroups within \texttt{SPRITZ}: AGN2, SB-AGN and SF-AGN galaxies. These subgroups are expected to have similar properties to QSO2, where dust obscuration plays a pivot role similarly to the \texttt{SPRITZ} subgroups, while having or not significant contribution from star-formation scenarios \citep[e.g.][see also references in Section \ref{section:SFR_separation}]{1984ApJ...282..427T,2003MNRAS.343..585F,2019MNRAS.489..427I,2020MNRAS.498.1560J}. 

\begin{figure*}
    \centering
    \includegraphics[width=0.7\linewidth]{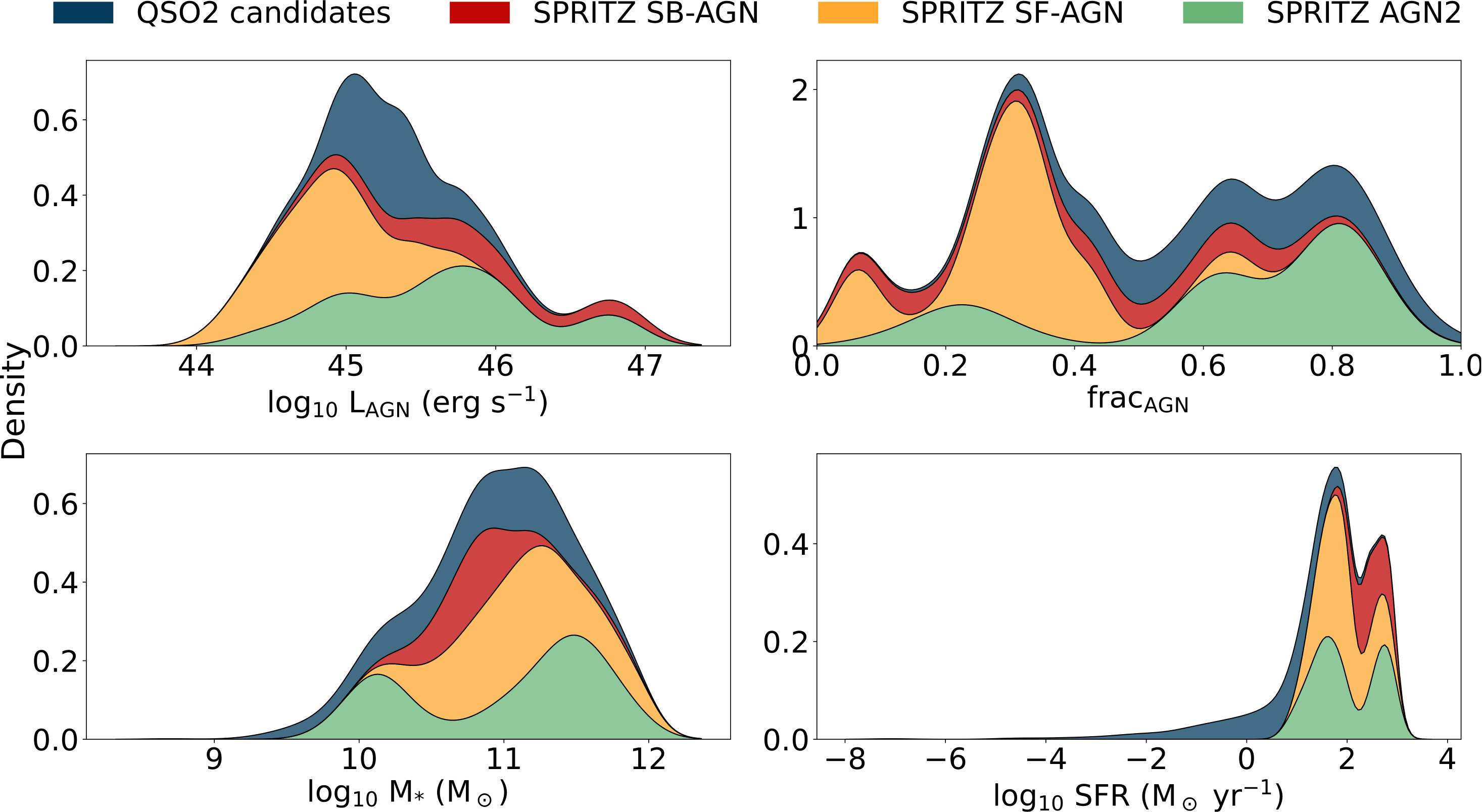}
    \caption{Distribution plots of the frac$_{\mathrm{AGN}}$ and L$_{\rm{AGN}}$ for the QSO2 candidates (blue), SB-AGN (red), SF-AGN (orange), and AGN2 (green) from \texttt{SPRITZ} catalogue.}
    \label{fig:compare_QSO2_SPRITZ}
\end{figure*}

To ensure the \texttt{SPRITZ} simulated data represents the distribution in the SDSS survey, we set a constraint on the number density of galaxies. We limited the number of galaxies within each infrared luminosity, i.e. L$_\text{IR}$-z redshift bin per square degree, to match the SDSS observations. The set the following threshold: N * SDSS area > 1, where N represents the simulated galaxy count. This criterion ensures that only galaxies common enough to be observed at least once within the SDSS area are included in our analysis. We then apply the same magnitude criteria as described in Section \ref{section:Data}. This step ensures the selection of optically faint sources, consistent with the selection criteria employed by both \citet{2013MNRAS.435.3306A} and our sample. Finally, we imposed constraints on the physical properties of the simulated galaxies, log$_{10}$(M$_{\ast}$/M$_\odot$) $ < 12$ and log$_{10}$(SFR [M$_\odot$ yr$^{-1}$]) $ < 3$. These limits are based on established physical boundaries for galaxies and filter out objects with unrealistically high stellar masses or star formation rates, which would create an unphysical bias.

In the \texttt{SPRITZ} catalogue, the physical properties were derived using the Multi-wavelength Analysis of Galaxy Physical Properties \citep[MAGPHYS]{2012IAUS..284..292D} code, or the \texttt{SED3FIT} \citep{2013A&A...551A.100B}, depending on the type of source and the presence of an AGN contribution. Although these codes differ from \texttt{CIGALE}, a comparison can be made to study the similarity between the simulated sources and our sample of QSO2 candidates.

In Figure \ref{fig:compare_QSO2_SPRITZ}, we show the density plots for the L$_{\rm{AGN}}$, frac$_{\mathrm{AGN}}$, M$_{\ast}$, SFR diagram for SF-AGN, SB-AGN, and AGN2 from \texttt{SPRITZ} alongside with our QSO2 candidates. The AGN sources selected from \texttt{SPRITZ} follow a similar distribution in the M$_{\ast}$-SFR diagram as the QSO2 candidates within and above the main-sequence, where the majority of candidates are placed. Although the estimates of physical properties used in SPRITZ and our work use different SED fitting codes, we observe a coherent range of physical properties between \texttt{SPRITZ} sources and our candidates.

Considering these results, we can argue that the QSO2 sample comprises a mix of composite systems, with both SB and SF-AGN galaxies, and more "traditional" AGN2 galaxies, whose host galaxy matches the properties of quiescent galaxies. To further study the true nature of our QSO2 sample, multi-wavelength analysis and morphological studies are necessary.

\section{Conclusions}
\label{section:conclusions}

We have studied the physical properties of a sample of QSO2 candidates in the redshift desert, i.e., $1 \leq z \leq 2$, selected via machine learning in C2024 \citep{2024A&A...687A.269C}, using the SED fitting code \texttt{CIGALE}. Our aim was to obtain new insights into galaxy and AGN evolution across cosmic time. 

We estimated relevant physical properties such as SFR, M$_{\ast}$, frac$_{\mathrm{AGN}}$, and AGN luminosity. Figure \ref{fig:SED_Mstellar_SFR} illustrates the distribution of SDSS galaxies (used as a control sample, i.e., galaxies classified as non-QSO2 candidates in C2024), QSO2 candidates, and the Bongiorno sample (a subset of \citet{2012MNRAS.427.3103B}) on the SFR-M$_{\ast}$ diagram. While SDSS galaxies are primarily concentrated in the high-SFR region, the QSO2 candidates exhibit a broader distribution, including sources with lower SFRs. The QSO2 candidates also span a wide range of SFRs and high M$_{\ast}$ values, from 10$^{9}$ to 10$^{12}$ M$_{\odot}$ (see Fig. \ref{fig:SED_delta_SFR}). We argue that our results demonstrate the ability of our methodology to identify QSO2 candidates independently of their host galaxy's physical properties. This can be helpful in unveiling new AGN2 candidates and increasing our knowledge and statistics on this type of AGNs.

Additionally, we computed the SFR/SFR$_{\rm{MS}}$ parameter, which allowed us to classify our candidates into four areas: starburst, log(SFR/SFR$_{MS}$) > 0.4 dex; star-forming, log(SFR/SFR$_{MS}$) $\pm$ 0.4 dex; sub-MS, log(SFR/SFR$_{MS}$) between -0.4 and -1.3; and quiescent, log(SFR/SFR$_{MS}$) < -1.3 \citep[e.g.,][]{2019MNRAS.484.4360A,2022A&A...659A.129V}. Our candidates are distributed across the four regions, with higher concentrations in the quiescent and star-forming regions.

Sources in the quiescent region have high M$_{\ast}$, significant frac$_{\mathrm{AGN}}$, and AGN luminosities matching the observed properties for AGN galaxies. The derived values for the AGN and host galaxy properties are consistent with those reported in the AGN literature \citep[e.g.,][]{Honig2007,2012MNRAS.427.3103B,2022A&A...659A.129V} and with those of Alexandroff sample \citep[L$_{\rm{AGN}} \sim$ 10$^{46.3-46.8}$ erg s$^{-1}$;][]{2013MNRAS.435.3306A}.

Speculating on their nature, we suggest that these sources may be subject to Compton-thick obscuration (N$_\mathrm{H} \sim$   10$^{24}$ cm$^{-2}$), which would lead to an underestimation of the SFR due to the significant attenuation of optical light. Alternatively, they could be experiencing the effects of AGN feedback. Further observational data, spanning from imaging to spectroscopy, are necessary to unveil their nature and the co-evolution of the AGN and host galaxy, as well as to test these hypotheses. In particular, far-infrared (FIR) data will be crucial to better constrain the physical properties derived in this study, especially SFR and AGN luminosities.

We also compared our QSO2 candidate sample with the semi-empirical simulated catalogue \texttt{SPRITZ}. We compared our sample with the composite systems (SB-AGN and SF-AGN) and the AGN2 sources. We observe a consistent distribution in the SFR-M$_{\ast}$ diagram when comparing with our QSO2 candidates. This result provides compelling evidence that supports the effectiveness of our methodology in selecting a diverse population of obscured AGNs.

In conclusion, we have studied the physical properties of our sample of QSO2 candidates in the redshift desert using the available optical and infrared photometry, from SDSS and WISE. The estimated physical properties and consequent analysis herein showcase the relevance of novel methodologies for the identification of QSO2 galaxies, particularly in the less explored redshift regime where studies are scarce. To further study the nature of QSO2 host galaxies, more multi-wavelength observations are essential \citep[e.g.,][]{2004AJ....128.1002Z,2005AJ....129.1212Z,2006ApJ...637..147P,2013MNRAS.436..997R}. For example, imaging observations would allow for morphological studies, which would help to study their merger history through disturbed morphologies or other distinctive features, testing the so-called merging paradigm \citep[e.g.,][]{2006AJ....132.1496Z,2011MNRAS.416..262V,2012MNRAS.426..276B,2012MNRAS.419..687R,2012MNRAS.423...80V,2013MNRAS.430.2327L,2015MNRAS.454.4452H,2015MNRAS.447.3322H,2016A&A...585A..32H,2023MNRAS.522.1736P}.


\begin{acknowledgements}
PACC thank the comments and suggestions from M. Villar-Mart{\'\i}n, R. Carvajal, and I. Matute.
PACC acknowledges financial support from Fundação para a Ciência e Tecnologia (FCT) through grant 2022.11477.BD, and through research grants UIDB/04434/2020 and UIDP/04434/2020. AH acknowledges support from NVIDIA through an NVIDIA Academic Hardware Grant Award.
      
This publication uses data products from the Wide-field Infrared Survey Explorer, which is a joint project of the University of California, Los Angeles, and the Jet Propulsion Laboratory/California Institute of Technology, funded by the National Aeronautics and Space Administration.

Funding for the Sloan Digital Sky Survey IV has been provided by the Alfred P. Sloan Foundation, the U.S. Department of Energy Office of Science, and the Participating Institutions. SDSS-IV acknowledges support and resources from the Center for High Performance Computing  at the University of Utah. The SDSS website is www.sdss.org. SDSS-IV is managed by the Astrophysical Research Consortium for the Participating Institutions of the SDSS Collaboration including the Brazilian Participation Group, the Carnegie Institution for Science, Carnegie Mellon University, Center for Astrophysics | Harvard \& Smithsonian, the Chilean Participation Group, the French Participation Group, Instituto de Astrof\'isica de Canarias, The Johns Hopkins University, Kavli Institute for the Physics and Mathematics of the 
Universe (IPMU) / University of Tokyo, the Korean Participation Group, Lawrence Berkeley National Laboratory, Leibniz Institut f\"ur Astrophysik Potsdam (AIP),  Max-Planck-Institut f\"ur Astronomie (MPIA Heidelberg), Max-Planck-Institut f\"ur Astrophysik (MPA Garching), Max-Planck-Institut f\"ur Extraterrestrische Physik (MPE), National Astronomical Observatories of China, New Mexico State University, New York University, University of Notre Dame, Observat\'ario Nacional / MCTI, The Ohio State University, Pennsylvania State University, Shanghai Astronomical Observatory, United Kingdom Participation Group, Universidad Nacional Aut\'onoma de M\'exico, University of Arizona, University of Colorado Boulder, University of Oxford, University of Portsmouth, University of Utah, University of Virginia, University of Washington, University of Wisconsin, Vanderbilt University, and Yale University.

In preparation for this work, we used the following codes for Python: Numpy \citet{harris2020array}, Scipy \citep{2020SciPy-NMeth}, Pandas \citep{mckinney-proc-scipy-2010}, matplotlib \citep{Hunter:2007}, and seaborn \citep{Waskom2021}.
\end{acknowledgements}

\section*{Data Availability}

The photometric data used were obtained through CasJobs\footnote{\url{https://casjobs.sdss.org/casjobs/}} and the SQL queries can be found here: \url{https://github.com/pedro-acunha/AMELIA/tree/main/data}. The machine learning code is available on GitHub: \url{https://github.com/pedro-acunha/AMELIA}. The derived physical properties of all candidates are available in electronic form at the CDS database.

\bibliographystyle{aa} 
\bibliography{QSO2_paperII} 

\appendix

\section{SED fitting parameters}

\begin{table*}
\centering
\caption{Summary of the \texttt{CIGALE} SED fitting parameters used for the control sample (galaxies).}
\begin{tabular}{@{}lcc@{}}
\hline 
Module & Parameter &  Values  \\
\hline 
Star formation history (\texttt{sfhdelayed}): & $e$-folding time $\tau_{\mathrm{main}}$ (Myr) & 5, 10, 30, 50, 100, 300, 500\\
 & Stellar Age, t$_{\mathrm{main}}$ (Myr)  & 15, 35, 40, 50, 75, 100, 200, 300, 500\\
 & $e$-folding time $\tau_{\mathrm{burst}}$ (Myr) & 5, 10, 15, 20 \\
 & Stellar Age, t$_{\mathrm{burst}}$ (Myr) & 2, 4, 6, 8, 10, 12, 14\\
 & Mass fraction, last burst & 0.0, 0.1, 0.3, 0.5 \\
\hline
Galactic dust emission (Dale et al. 2014): & $\alpha$ slope & 0.5, 1, 1.5, 2.0, 2.5, 3, 3.5\\
 & frac$_{\mathrm{AGN}}$ & 0.0, 0.1, 0.25\\
\hline
\texttt{dustatt\_modified\_starburst}: & E(B-V) & 0.1, 0.2, 0.4, 0.6, 0.7, 0.8, 0.9\\
\hline 
\end{tabular}
\label{tab:SED_fitting_params_gal}
\end{table*}

\begin{table*}
\begin{center}
\caption{Summary of the \texttt{CIGALE} SED fitting parameters used for the QSO2 candidates.}
\begin{tabular}{@{}*{3}{c}@{}}
\hline 
Module & Parameter &  Values  \\
\hline 
 Star formation history: & $e$-folding time $\tau_{\rm{main}}$ (Myr) & 5-20 (step 5), 30, 50, 100-300 (step 100),\\
 & &  500.0, 1000.0, 3000.0, 5000.0 \\
 \texttt{sfhdelayed} & Stellar Age, t$_{\rm{main}}$ (Myr)  & 31, 35, 40, 50, 75, 100-500 (step 100), \\
 & & 1000, 3000, 4000, 5000, 5500\\
 & $e$-folding time $\tau_{\rm{burst}}$ (Myr) & 5, 10, 15, 20, 25, 50 \\
 &  Stellar Age, t$_{\rm{burst}}$ (Myr)& 5, 10, 15, 20, 25, 30\\
 & Mass fraction, last burst & 0.0, 0.1, 0.2 \\
 \hline
 Galactic dust emission (Dale et al. 2014): & $\alpha$ slope & 0.5, 1, 1.5, 2.0, 2.5, 3, 3.5\\
\hline 
 AGN: \texttt{SKIRTOR}  & Average edge-on optical depth at 9.7 $\mu$m & 3, 7, 11 \\
 & AGN contribution to IR luminosity, frac$_{\mathrm{AGN}}$ & 0.0-0.9 (step 0.1), 0.99\\
  Stalevski et al. (2016) & inclination (i.e. viewing angle) & 40, 60, 80\\
  & Intrinsic disk type & Schartmann (2005) spectrum \\
  & Polar extinction E(B-V) &  0, 0.05, 0.1, 0.2, 0.3 \\
\hline 
\end{tabular}
\label{tab:SED_fitting_params_QSO2}
\end{center}
\end{table*}

\section{\texttt{LePhare++} photo-z estimations}
\label{appendix_LePhare}
\begin{figure*}
    \centering
    \includegraphics[width=\linewidth]{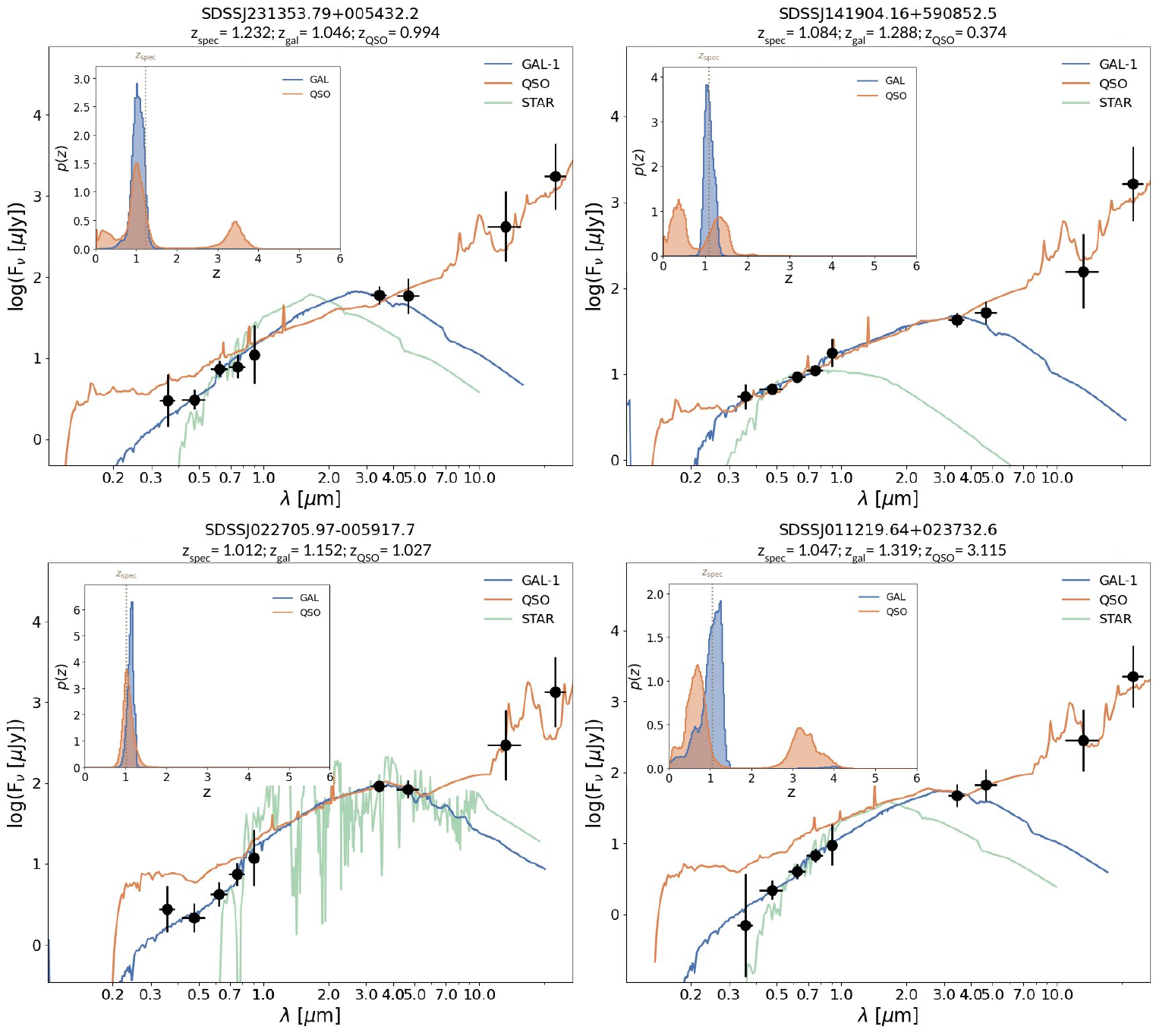}
    \caption{Examples of the best-fitting models for photometric redshift estimation using \texttt{LePhare++} for four sources. Filled symbols represent the observed flux densities from SDSS and WISE. Source classification is defined by the reduced $\chi^2$ value. The best-fit galaxy template is shown in blue, the best-fit QSO template in orange, and the best-fit star template in green. The histograms within each plot show the probability distribution function for the photometric redshift estimation, with the vertical dashed line indicating the spectroscopic redshift from SDSS.}
    \label{fig:LePhare_examples}
\end{figure*}

We present examples of photometric redshift estimation using \texttt{LePhare++}. By fitting templates to observed flux densities from SDSS and WISE, we classify sources based on the reduced $\chi^2$ value, as shown in Figure \ref{fig:LePhare_examples}. The fitting process includes galaxy, QSO, and star templates, highlighting the challenges \texttt{LePhare++} faces with composite systems. In such cases, the optical portion of the SED often aligns closely with the galaxy template, while the IR portion is dominated by the QSO template. These examples show the limitations of current templates in accurately modelling optical-obscured AGNs.

\section{\texttt{CIGALE} fitting examples}
\begin{figure*}
    \centering
    \includegraphics[width=\linewidth]{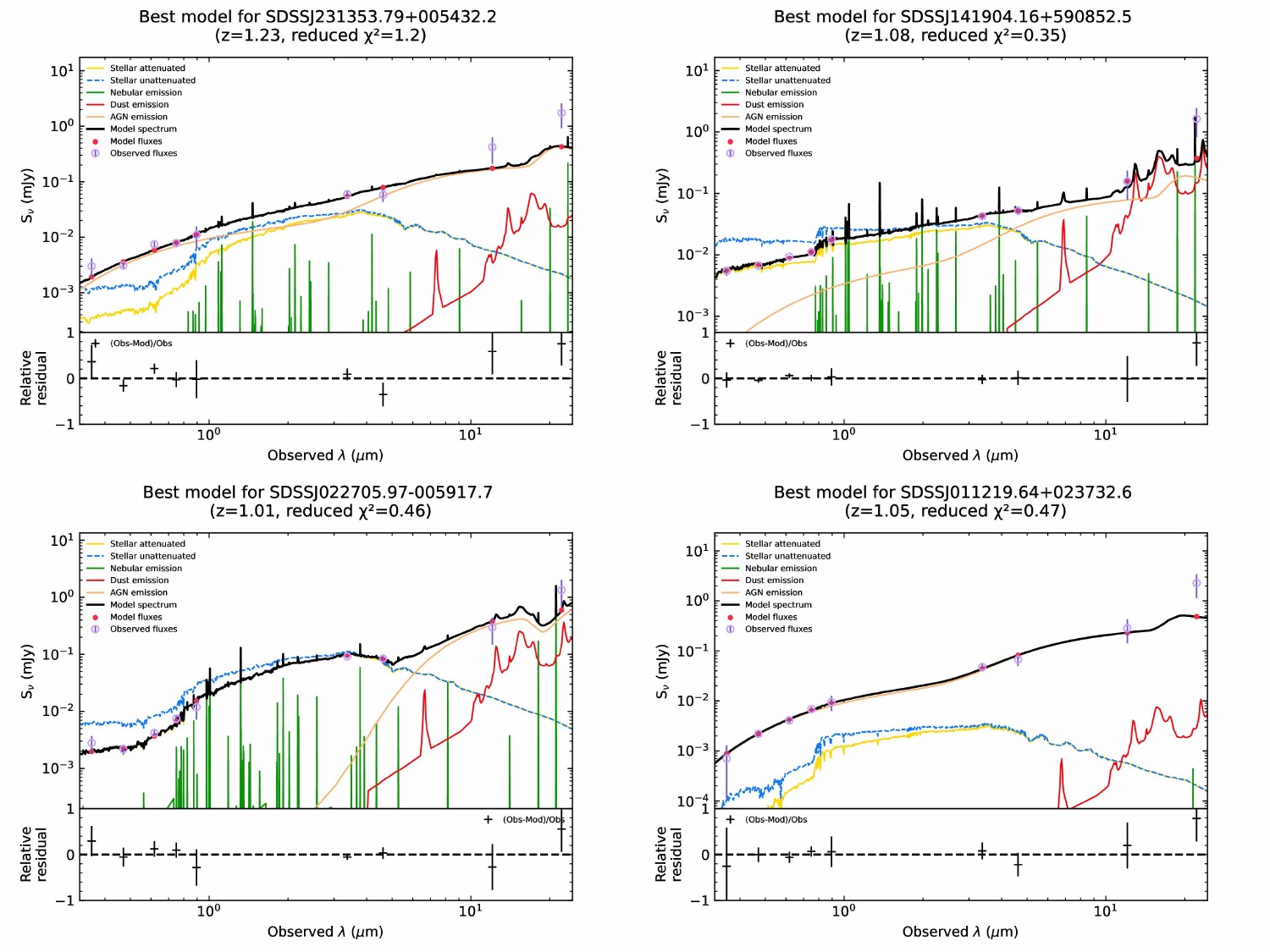}
    \caption{Best-fitting models for four candidates using \texttt{CIGALE}, similar to Figure \ref{fig:LePhare_examples}. Open symbols represent observed flux densities, while the red filled symbols indicate modelled flux densities. The goodness of fit is shown by the reduced $\chi^2$ value, with the residuals of the fit displayed at the bottom of each panel. The derived physical properties are shown in Table \ref{tab:SED_example}.}
    \label{fig:SED_example}
\end{figure*}

In this section, we present four example fits generated with \texttt{CIGALE}, corresponding to the same candidates shown in Section \ref{appendix_LePhare}. Each plot illustrates the contributions of various components to the overall SED of each candidate: the attenuated stellar emission is shown in yellow, unattenuated stellar emission in blue, nebular emission in green, dust emission in red, and AGN emission in orange. The complete model spectrum, combining all components, is represented by the black line. These examples show how more complex models, such as \texttt{SKIRTOR}, are important to model the AGN contribution to the SED of our candidates. For comparison purposes, we show in Table \ref{tab:SED_example} the physical properties derived for the sources shown in Figure \ref{fig:SED_example}.

\begin{table}[H]
    \caption{Physical properties derived using \texttt{CIGALE} for the four sources shown in Figures \ref{fig:LePhare_examples} and \ref{fig:SED_example}.} 
    \centering
    \begin{tabular}{c c c c c}
    \hline
        ID & frac$_{\mathrm{AGN}}$ & M$_{\ast}$ & SFR & L$_{\rm{AGN}}$\\
    \hline
        SDSSJ231353.79+005432.2 & 0.88 & 10.12 & 1.08 & 38.55 \\
        SDSSJ141904.16+590852.5 & 0.63 & 10.07 & 1.31 & 38.18 \\
        SDSSJ022705.97-005917.7 & 0.63 & 11.27 & 0.89 & 38.15 \\
        SDSSJ011219.64+023732.6 & 0.97 & 9.62 & 0.09 & 38.33 \\
    \hline
    \end{tabular}
    \label{tab:SED_example}
    \tablefoot{All physical properties, with the exception of frac$_{\mathrm{AGN}}$, are presented in logarithmic scale.}
\end{table}

\section{SED mock comparison}
\label{appendix:SED_mock}
To understand whether the estimated physical properties using \texttt{CIGALE}, with SDSS and WISE photometry, we used the mock catalogue option provided in the \texttt{CIGALE} configuration file. 

\begin{figure*}
    \centering
    \includegraphics[width=0.8\linewidth]{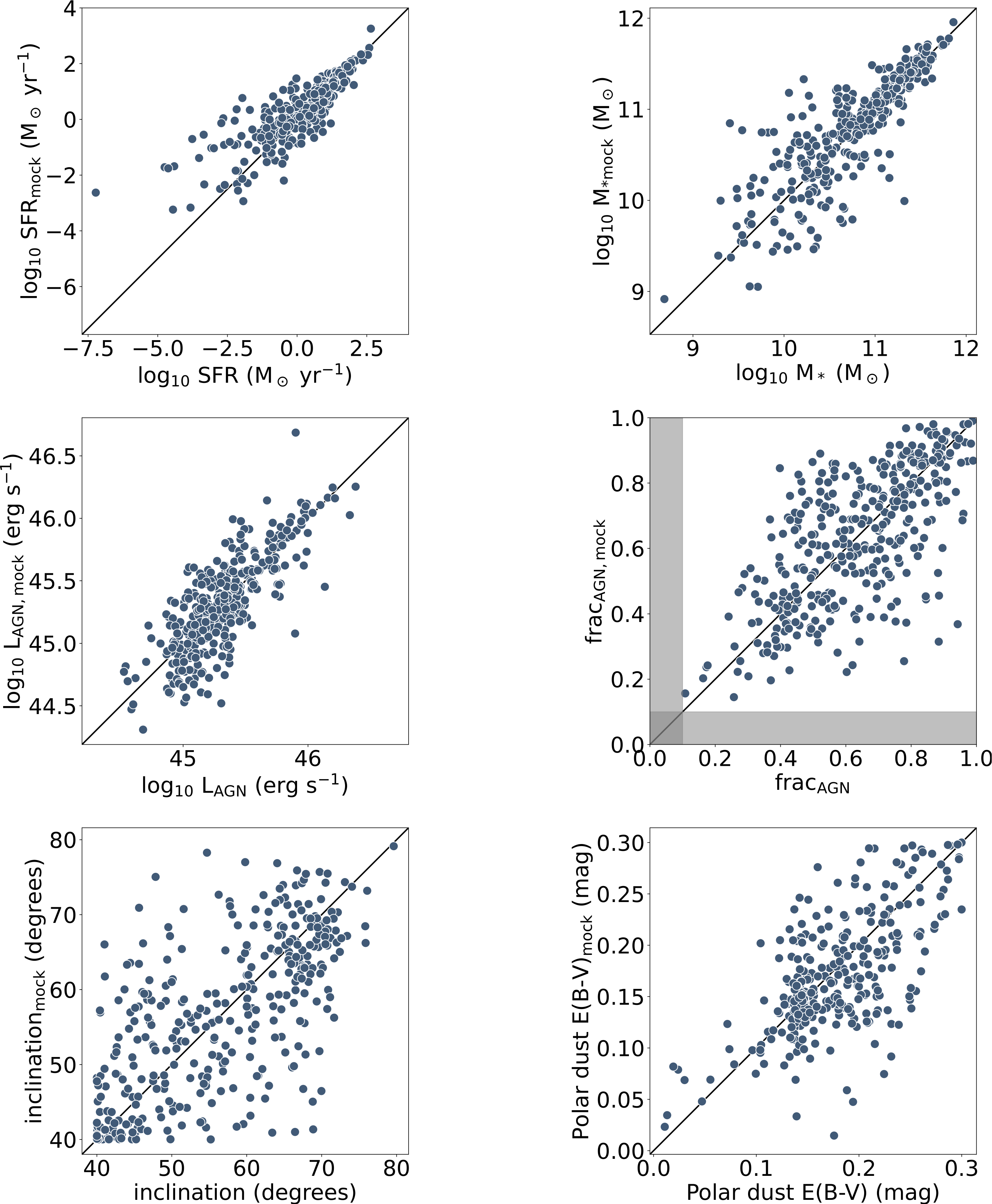}
    \caption{Comparison between the estimated physical properties as derived from the mock analysis and the estimated values using \texttt{CIGALE}. The grey dashed line indicates the 1:1 relation between the parameters. From top to bottom, left to right: star-formation rate; stellar mass; AGN luminosity; AGN fraction; viewing angle (inclination); and polar-dust extinction, E(B-V).}
    \label{fig:mock_analysis}
\end{figure*}

The mock catalogue is built, for each source, using the best fit and introducing random errors to each quantity from a Gaussian distribution with the same standard deviation as the observation uncertainty. In Figures \ref{fig:mock_analysis}, we show the linear correlation between the mock and predicted values.

\end{document}